\PassOptionsToPackage{bookmarks=false}{hyperref}
\documentclass[sn-mathphys-num,Numbered]{sn-jnl}
\usepackage{amsmath,amssymb,amsfonts}
\usepackage{graphicx}
\usepackage{textcomp}
\usepackage{xcolor}
\usepackage{enumitem}
\usepackage{booktabs}
\usepackage{multirow}
\usepackage{natbib} 

\newcommand{\approach}{{CIgrate}}
\newcommand{\prevapproach}{\text{CIMig}}
\newcommand{\summarybox}[1]{
    \vspace{3mm}
    \noindent 
    \framebox[\linewidth][c]{\parbox[b]{0.95\linewidth}{{#1}}}
}

\usepackage{caption}
\usepackage{array}
\usepackage{ragged2e}
\newcolumntype{L}[1]{>{\raggedright\let\newline\\\arraybackslash\hspace{0pt}}m{#1}}
\newcolumntype{C}[1]{>{\centering\let\newline\\\arraybackslash\hspace{0pt}}m{#1}}
\newcolumntype{R}[1]{>{\raggedleft\let\newline\\\arraybackslash\hspace{0pt}}m{#1}}

\usepackage{listings}

\lstdefinestyle{yml}{
     float=tp,
     floatplacement=tbp,
     abovecaptionskip=-5pt,
     numberstyle=\normalfont\tiny\color{gray},
     basicstyle=\color{black}\footnotesize\ttfamily,
     rulecolor=\color{black},
     string=[s]{'}{'},
     keywordstyle=\color{red},
     morecomment=[l]{-},
     morecomment=[l]{+},
     moredelim=[is][\color{red}]{|<}{>|},
     moredelim=[is][\color{blue}]{|>}{<|},
}

\usepackage[most]{tcolorbox}
\definecolor{DarkGreen}{RGB}{1,150,32}
\lstdefinestyle{PromptStyle}{
  basicstyle=\ttfamily\scriptsize,
  breaklines=true,
  frame=single,
  captionpos=b,
  rulecolor=\color{gray},
  escapeinside={(*@}{@*)},
  moredelim=**[is][\color{DarkGreen}\bfseries]{@d@}{@},
  moredelim=**[is][\color{blue}\bfseries]{@u@}{@},
  moredelim=**[is][\color{black}\bfseries]{@s@}{@}
}

\begin{document}

\title{CIgrate: Automating CI Service Migration with Large Language Models
}

\author[1]{\fnm{Md Nazmul} \sur{Hossain}}\email{nazmulhossain@trentu.ca}

\author[1]{\fnm{Taher A.} \sur{Ghaleb}}\email{taherghaleb@trentu.ca}

\affil[1]{\orgname{Trent University}, \city{Peterborough}, \country{Canada}}

\abstract{
Continuous Integration (CI) configurations often need to be migrated between services (e.g., Travis CI to GitHub Actions) as projects evolve due to changes in service capabilities, usage limits, or service deprecation. Previous studies reported that migration across CI services is a recurring need in open-source development. However, manual migration is time-consuming and error-prone. The state-of-the-art approach, {\prevapproach}, addresses this challenge by analyzing past migration examples to create service-specific rules and produce equivalent configurations across CI services. However, its relatively low accuracy (0.49 Cosine Similarity for Travis$\rightarrow$GHA) raises concerns about the feasibility of rule-based CI migration alone. Meanwhile, Large Language Models (LLMs) have demonstrated strong capabilities in code generation and transformation tasks, suggesting potential to improve the automation, usability, and generalizability of CI configuration migration.
This paper presents an empirical study assessing whether CI migration can be improved using LLMs. We propose {\approach}, an LLM-based framework for automatically migrating CI configurations, and compare it with {\prevapproach} using (a) zero-shot/few-shot prompting and (b) fine-tuning on a dataset of established CI migrations. We also evaluate practical deployment by submitting pull requests to active open-source projects. Our results show that {\approach} substantially outperforms {\prevapproach}: the fine-tuned Gemma~3 12B achieves 0.90 Cosine Similarity and 0.74 CrystalBLEU for Travis CI (Travis)$\rightarrow$GitHub Actions (GHA) migrations (+82.2\% and +295.5\% over {\prevapproach}), while producing syntactically valid, immediately parseable YAML in 100\% of cases, compared to 5.6\% for {\prevapproach}. Even zero-shot LLMs outperform the rule-based baseline, demonstrating that LLM-based approaches provide a more practical and accurate solution for CI configuration migration.
}


\maketitle

\section{Introduction}
\label{intro}

Continuous Integration (CI) configurations define how a CI service automates software builds, tests, and deployments~\cite{Fowler_CI}. As projects evolve, developers often need to migrate their CI configurations between services, such as Travis CI to GitHub Actions~\cite{Hilton2016,Zampetti2019,Zhao2017}. Prior studies show that developers often switch between CI services due to evolving project needs, tool limitations, or service deprecation~\cite{woodward2020travis,rzig2022characterizing}. 
Though the frequency of such migrations is not broadly quantified, previous studies have shown that they are a recurring need in open-source development~\cite{rostami2023usage}.
Rzig et al.~\cite{rzig2024cimig} highlighted the challenges of manual migration as being time-consuming and error-prone, underscoring the need for automated solutions. GitHub Actions, introduced in 2019 as an integrated CI solution~\cite{githubactions}, has since become widely adopted, while Travis CI has declined in popularity due to service limitations and usage limits~\cite{woodward2020travis}. Nevertheless, some projects may still require Travis CI support as a backup CI service or for compatibility with specific infrastructure. CI pipeline migration helps improve matrix support, containerization, and seamless integration workflows. However, manually rewriting configuration files is slow and can lead to mistakes~\cite{rostami2023usage}. CI configurations are also increasingly complex and often lack standardization across services, making automated migration non-trivial~\cite{ghaleb2024cicd}.

Rzig et al.~\cite{rzig2024cimig} proposed {\prevapproach}, an automated approach for migrating CI configurations using example-based rule mining, enabling seamless transitions between services like Travis CI and GitHub. Their approach improves migration coverage and reduces developer effort by over 40 minutes~\cite{rzig2024cimig}. It migrates CI configurations between Travis CI and GitHub Actions, relying on past examples to generate equivalent configurations across CI services. However, it faces challenges with complex job matrices and custom environment variables, and requires access to high-quality migration pairs for more accurate results~\cite{rzig2024cimig}. Besides, its relatively low accuracy (achieving only 0.49 Cosine Similarity and producing syntactically valid YAML in only 5.6\% of cases) raises concerns about the overall feasibility of automated CI migration using rule-based techniques alone, as rule-based approaches struggle with semantic preservation when CI features lack direct structural equivalents across services.

LLMs have demonstrated strong capabilities in code generation and transformation tasks~\cite{jiang2024survey,chen2024survey}.
To ensure a comprehensive and rigorous evaluation, {\approach} adopts a diverse set of LLMs, including three open-weight models: Gemma~3~\cite{gemma2024}, Llama~3.1~\cite{llama2024}, and Mistral~0.3~\cite{mistral2023}, informed by a prior study~\cite{wang2025llm}. We also include a proprietary LLM (GPT-4o), widely recognized as a benchmark model in academia and industry for its coding and reasoning capabilities~\cite{openai2024gpt4o,zheng2025towards}. This balanced combination enables us to rigorously evaluate {\approach} against both proprietary and accessible open-weight LLMs, capturing a realistic range of model capabilities and performance.

However, to our knowledge, no previous work has systematically evaluated the effectiveness of LLMs for CI migration tasks compared to {\prevapproach}.
This paper presents an empirical study in which we propose {\approach}, a framework for automatically migrating CI configurations between Travis CI and GitHub actions, and vice versa. Unlike {\prevapproach}, which depends on predefined rules or examples, {\approach} employs LLMs to enable two-way migration with minimal effort.
We perform multiple empirical analyses to a) measure the accuracy of LLM-generated outputs in both zero-shot, few-shot, and fine-tuned settings, 2) compare LLM-generated outputs with those of the state-of-the-art approach ({\prevapproach}) using similarity metrics, and 3) evaluate practical deployment by submitting pull requests to active projects and measuring CI execution success rates. By integrating quantitative analyses (e.g., similarity metrics) with qualitative assessments (i.e., developer feedback), and combining automated evaluation techniques with human judgment, our study adopts an empirical approach that offers actionable insights for both researchers and practitioners in the software engineering community.

To achieve this, we use a curated dataset from Rzig et al.~\cite{rzig2024cimig}, containing real CI migrations between Travis CI and GitHub Actions. Our evaluation of {\approach} integrates automated metrics (e.g., similarity and accuracy) with developer feedback to assess its accuracy and usability. This mixed-method approach helps to understand LLM performance in real migration scenarios and the trade-offs involved. Our findings help CI practitioners design future LLM-based solutions and integrate them into CI workflows.

The rest of this paper is organized as follows.
Section \ref{sec:background_and_related_work} provides a background context and an overview of the related work.
Section~\ref{sec:approach} describes our automated CI migration approach.
Section~\ref{sec:evaluation} presents the experimental design.
Section~\ref{sec:results} reports the results of our evaluation.
Section~\ref{sec:discussion} discusses the practical implications of our findings.
Section~\ref{sec:threats_to_validity} discusses the validity threats to our results. Section~\ref{sec:conclusion} draws conclusions and suggests future work.

\section{Background and Related Work}
\label{sec:background_and_related_work}

\subsection{Background}
\begin{figure}[ht] 
    \centering
    \vspace{-8pt}
    \includegraphics[width=0.9\textwidth, trim=18 0 0 0, clip]{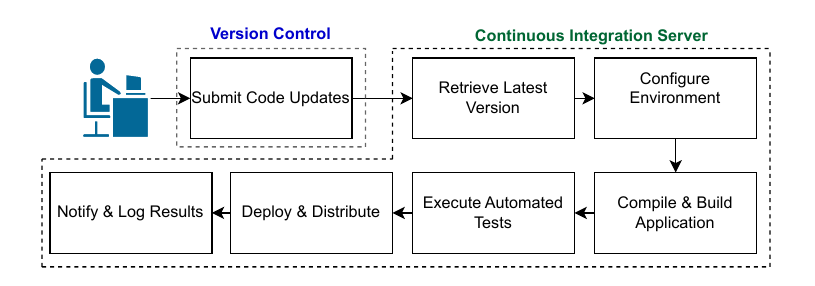}
    \caption{CI build lifecycle}
    \vspace{-5pt}
    \label{fig:ci_lifecycle} 
\end{figure}

Continuous Integration (CI) configuration files define how source code is built, tested, and deployed in automated pipelines~\cite{duvall2007continuous}. CI follows a well-defined life cycle for generating builds. Figure \ref{fig:ci_lifecycle} illustrates the main phases of the CI build life cycle. CI services such as Travis CI~\cite{TravisCI2022} and GitHub Actions~\cite{githubactions} require developers to express this logic using YAML-based~\cite{travisci2021yaml} configuration files~\cite{belmont2018hands,heller2021automating}. For example, a typical `\texttt{.travis.yml}' file in Travis CI might declare the programming language, build matrix, script commands, and deployment steps~\cite{belmont2018hands}. In GitHub Actions, these are written in `\texttt{.yml}' workflow files placed under the `\texttt{.github/workflows}' directory, with a more structured syntax based on jobs, steps, triggers, and reusable actions~\cite{heller2021automating}.

Though both formats use YAML~\cite{travisci2021yaml}, their schemas and semantics differ significantly. For instance, Travis CI specifies jobs via language directives and matrix expansions, whereas GitHub Actions requires explicit job declarations with `\texttt{runs-on}' keys and step sequences~\cite{githubactions,travisci}. Even minor differences, such as the handling of environment variables or conditional job execution, can cause functional inconsistencies after migration. Moreover, some Travis features (e.g., `\texttt{addons}' or `\texttt{sudo}') do not have direct equivalents in GitHub Actions, making one-to-one migration difficult~\cite{githubactions,travisci}.

\begin{figure}[t]
    \centering
    \includegraphics[width=\textwidth]{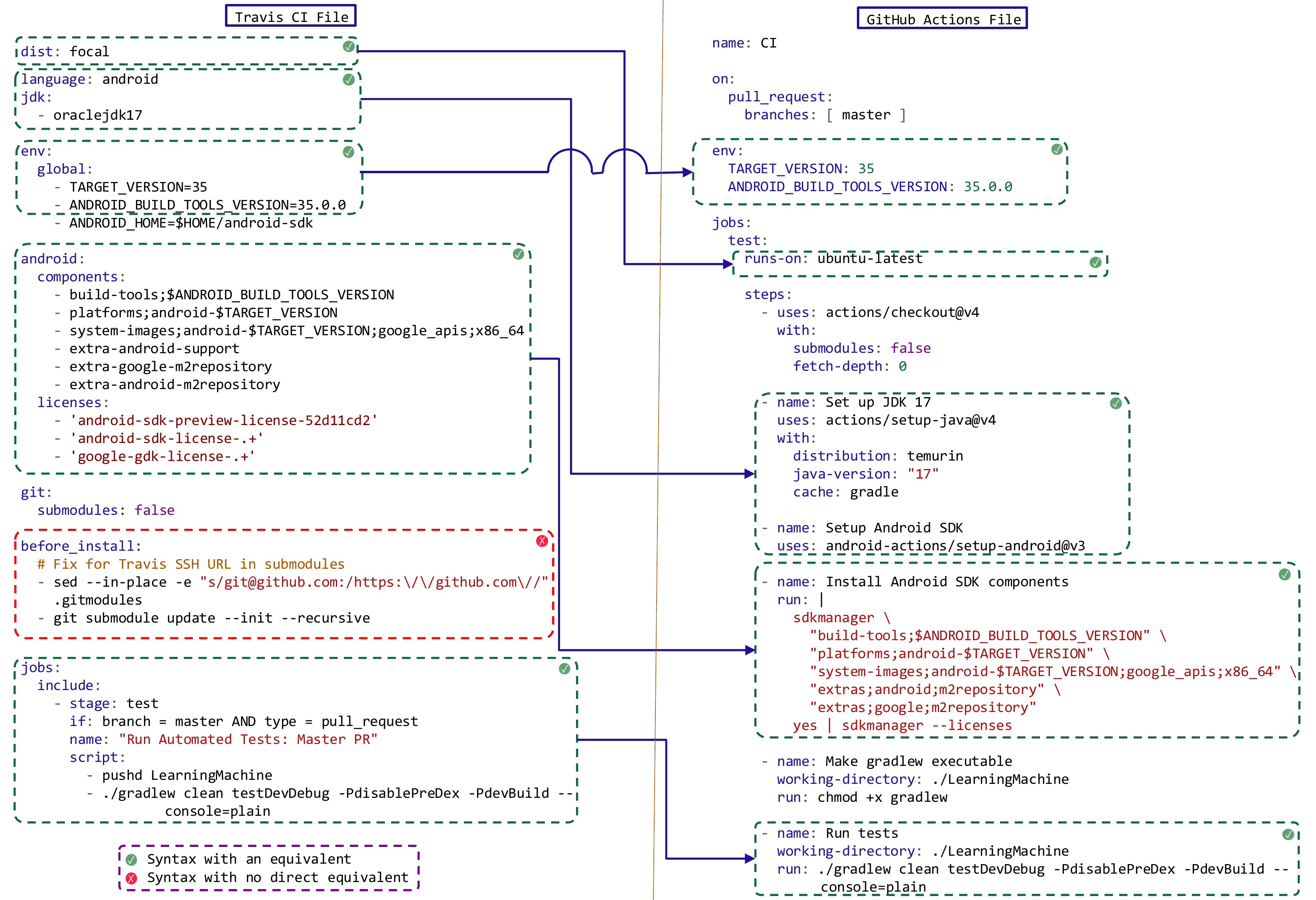}
    \caption{Example Configurations of Travis CI and GitHub Actions. {\color{green} $\checkmark$} means syntax with equivalence, {\color{red} $\times$} means syntax with no direct equivalence}
    \label{fig:motivational_example}
\end{figure}

To illustrate the complexity of CI configuration migration, Figure~\ref{fig:motivational_example} presents an example migration from Travis CI to GitHub Actions for an Android project.

The example demonstrates several non-trivial mapping challenges that automated migration tools must handle. First, Travis CI's \texttt{env.global} variables map directly to GitHub Actions' top-level \texttt{env} block, but the syntax differs (\texttt{- KEY=value} vs.\ \texttt{KEY: value}). Second, Travis CI's \texttt{language: android} with \texttt{jdk: oraclejdk17} must be translated to explicit setup actions (\texttt{actions/setup-java@v4} with \texttt{distribution: temurin}) since GitHub Actions does not have native language directives. Third, Travis CI's \texttt{android.components} block, which declaratively specifies SDK components to install, must be converted to imperative \texttt{sdkmanager} commands within a run step. Fourth, the \texttt{git.submodules: false} setting maps to the \texttt{submodules} parameter of \texttt{actions/checkout@v4}, and Travis's \texttt{before\_install} hook for submodule initialization is no longer needed since the checkout action handles this natively. Finally, Travis CI's job matrix with conditional execution (\texttt{if: branch = master AND type = pull\_request}) maps to GitHub Actions' \texttt{on.pull\_request.branches} trigger, with the script commands becoming discrete steps with explicit \texttt{working-directory} specifications.

{\prevapproach}~\cite{rzig2024cimig} is a tool that enables automating the migration of the CI configurations using a learning-based approach. It extracts transformation rules from aligned examples of `\texttt{.travis.yml}' and GitHub Actions workflows across thousands of open-source repositories. {\prevapproach} then applies association rule mining and abstract syntax tree (AST) matching to translate configurations. The tool supports two-way migration and was shown to reduce manual migration time by more than 42 minutes per project on average. We use {\prevapproach} as a baseline to compare with our proposed framework.

\subsection{Related Work}
The problem of migrating CI configurations intersects with broader efforts in software maintenance and automated transformation~\cite{rostami2023usage}. Prior research has shown that CI configuration issues, such as misused caching, retry logic, or incomplete setups, can lead to longer build times and failures~\cite{ghaleb2019duration,ghaleb2019noise} and conflicting trade-offs between speed and reliability~\cite{ghaleb2022interplay}. Another study showed that developers tend to abandon using CI due to many issues, including configuration~\cite{widder2018m}. Recent studies have also examined the broader landscape of CI/CD automation, analyzing how automated pipelines enhance developer productivity~\cite{chittala2024enhancing} and identifying research gaps in CI/CD practices~\cite{baitha2024streamlining}.

Rostami et al.~{\cite{rostami2023usage}} showed that developers sometimes adopt more than one CI service simultaneously, suggesting a trend of transitioning between CI services. For example, after the release of GitHub Actions, many projects migrated away from other services such as Travis CI, highlighting the need for automated support to facilitate such migrations. Conversely, developers may also need to add Travis CI as a backup CI service, motivating migration in the opposite direction. To address this need, Rzig et al.~\cite{rzig2024cimig} proposed a technique called \prevapproach, which supports migration between Travis CI and GitHub Actions in both directions. {\prevapproach} relies on examples collected from GitHub repositories using both services and applies rule generalization to derive flexible mappings. It represents the state-of-the-art technique on automated CI service migration.

Recent advances in LLMs have influenced software engineering tasks, including code completion~\cite{zheng2025towards,hou2024large,wang2024software}. These models, pre-trained on massive amounts of data, are capable of code generation, repair, and even translating source code across programming languages~\cite{yuan2024transagent, yang2024exploring}. Code-to-code translation has received growing attention, with recent work exploring error correction methods~\cite{xue2024interpretable} and few-shot learning approaches~\cite{li2024few} to improve translation accuracy. Building on these capabilities, LLMs show potential for synthesizing CI configurations and translating them across different formats without relying on explicit rules. Though LLMs have been increasingly applied to software development tasks, no prior work has empirically evaluated an end-to-end LLM-based solution for CI configuration migration. This leaves an open research gap in assessing the effectiveness and usability of LLM-powered migration solutions compared to rule-based and example-driven approaches in real-world settings.

This research addresses the gap by introducing and benchmarking {\approach}, an LLM-based migration framework that employs multiple models under different settings (zero-shot, few-shot, and fine-tuned) to automate CI service transitions. We evaluate {\approach} under various configurations and compare it against {\prevapproach} using real-world, two-way migration tasks between Travis CI and GitHub Actions. The study also includes developer feedback to assess the perceived quality, usability, and trustworthiness of the outputs generated by our framework.

\section{\approach: LLM-based Automated CI Service Migration}
\label{sec:approach}

This section presents {\approach}, a framework that uses Large Language Models (LLMs) to migrate CI configurations between Travis CI and GitHub Actions in both directions. {\approach} is designed to (1) preserve the intent and key features of the source CI pipeline, (2) generate syntactically valid YAML in the target CI format, and (3) make limitations explicit when a source feature has no direct equivalent in the target service.

\subsection{Overview}
{\approach} takes as input a CI configuration file from a source service (either \texttt{.travis.yml} or a GitHub Actions workflow \texttt{.yml}) and produces a target configuration in the other service’s format. The core of {\approach} is an instruction-driven LLM prompt that enforces strict output constraints: the model must output only YAML, follow the target service syntax, preserve semantic intent, and add a \texttt{\# TODO} YAML comment only when a feature cannot be faithfully mapped.

Figure~\ref{fig:study_overview} (Section~\ref{sec:evaluation}) illustrates the overall workflow of our study using {\approach} across different prompting modes and model setups.

\subsection{Language Models Used}
\label{sec:models_used}

We evaluate {\approach} using a mix of proprietary and open-weight LLMs to cover a realistic range of model capabilities and deployment settings:
\begin{itemize}[leftmargin=*]
    \item \textbf{GPT-4o} (proprietary model), used for zero-shot and few-shot.
    \item \textbf{Gemma 3 (12B)} (open-weight model), used for zero-shot, few-shot, and fine-tuning.
    \item \textbf{Mistral 0.3 (7B)} (open-weight model), used for zero-shot and few-shot.
    \item \textbf{Llama 3.1 (8B)} (open-weight model), used for zero-shot and few-shot.
\end{itemize}

All models are used with deterministic decoding by setting temperature to 0, which matches the practical usage scenario where developers typically expect a single migration output rather than sampling multiple candidates.

We use GPT-4o as the proprietary baseline because it was OpenAI's flagship model at the time the experiments were conducted. GPT-4o provides state-of-the-art coding and reasoning capabilities relevant to CI migration~\cite{openai2024gpt4o}.

\subsection{Two-Way Migration}
\label{sec:two_way_migration}

{\approach} supports two migration directions:

\paragraph{Travis CI to GitHub Actions.}
Given a \texttt{.travis.yml}, {\approach} generates a GitHub Actions workflow file under \texttt{.github/workflows/}, preserving stages such as environment setup, dependency installation, caching, build, test, and deployment.

\paragraph{GitHub Actions to Travis CI.}
Given a GitHub Actions workflow, {\approach} generates an equivalent Travis CI configuration, mapping jobs and steps into Travis lifecycle phases (\texttt{before\_install}, \texttt{install}, \texttt{script}, \texttt{after\_success}, \texttt{deploy}, etc.) and translating matrices into Travis build matrix syntax where applicable.

When an exact mapping is not possible (for example, certain GitHub Actions triggers or reusable workflows), {\approach} inserts a YAML comment in the output to flag the limitation for manual resolution.

\subsection{Prompting Modes}
\label{sec:prompting_modes}

{\approach} is evaluated in three modes: zero-shot, few-shot, and fine-tuned. The prompting design uses a generic system and user template with direction-specific placeholders and guideline blocks to enforce syntax correctness and semantic intent.

\subsubsection{System and User Prompts}
\label{sec:prompts}

CI configuration migration is challenging because it requires simultaneously (i) producing \emph{valid target syntax} that conforms to the destination CI schema and (ii) preserving the \emph{intended pipeline semantics} (e.g., triggers, environment setup, build/test steps, and deployment conditions). To support consistent evaluation across models and prompting modes, we adopt a standardized two-part prompting design: a \textbf{generic system prompt template} that establishes the model role and global constraints, and a \textbf{generic user prompt template} that specifies the migration task, injects direction-specific guidelines, and embeds optional in-context examples together with the input configuration.

\paragraph{System prompt.}
We use a single \emph{generic} system prompt template for both migration directions. The system message frames the model as an expert DevOps engineer experienced with both GitHub Actions and Travis CI, and parametrizes the task using placeholders for \textit{\{\{ SOURCE\_CI \}\}} and \textit{\{\{ TARGET\_CI \}\}}. It then enforces a fixed set of \emph{core requirements}: the model must output \emph{only} pure YAML (no explanations, markdown, or additional text), comply strictly with \textit{\{\{ TARGET\_CI \}\}} syntax, preserve semantic equivalence with the \textit{\{\{ SOURCE\_CI \}\}} pipeline, follow \textit{\{\{ TARGET\_CI \}\}} best practices, and avoid brittle line-by-line translation in favor of reproducing the overall pipeline behavior in the target workflow.

\paragraph{User prompt template.}
The user prompt is likewise defined as a generic template shared across both directions. It (i) states the migration request (migrating \textit{\{\{ SOURCE\_CI \}\}} to \textit{\{\{ TARGET\_CI \}\}}, (ii) reiterates hard requirements such as strict \textit{\{\{ TARGET\_CI \}\}} YAML compliance and functional preservation, and (iii) includes a direction-specific guideline block (instantiated depending on the migration direction). The template also requires the model to flag any \textit{\{\{ SOURCE\_CI \}\}} feature lacking a direct \textit{\{\{ TARGET\_CI \}\}} equivalent using an inline YAML comment (e.g., \texttt{\# TODO: ...}) for manual resolution.

The direction-specific guideline block provides tailored instructions depending on whether the migration is Travis~CI$\rightarrow$GitHub~Actions or vice versa. For Travis~CI$\rightarrow$GitHub~Actions, the guidelines instruct the model to use appropriate GitHub Actions syntax (e.g., \texttt{runs-on} for specifying runners, \texttt{uses} for invoking marketplace actions, and \texttt{on} for defining triggers), and to translate Travis lifecycle phases (\texttt{before\_install}, \texttt{install}, \texttt{script}, etc.) into discrete workflow steps. For GitHub~Actions$\rightarrow$Travis~CI, the guidelines direct the model to map explicit job and step declarations into Travis's phase-based structure, convert action invocations to equivalent shell commands or Travis addons where applicable, and express build matrices using Travis's \texttt{env} and \texttt{matrix} syntax.

To support few-shot prompting, the template includes a placeholder field \textit{\{\{ FEW\_SHOT\_EXAMPLES \}\}}, which is empty in zero-shot mode and populated with three example migrations in few-shot mode. Finally, the source configuration to migrate is injected via \mbox{\textit{\{\{ INPUT\_CODE \}\}}}.

Listing~\ref{lst:system-user-prompt-template} shows the full prompt template used in {\approach}, highlighting the shared skeleton and the placeholders used to instantiate the migration direction, optional few-shot examples, and the input configuration. This standardized design ensures all models are evaluated under the same prompt structure, isolating the effects of prompting mode and model choice.

The prompt template was developed through multiple iterations of piloting and refinement. Initial versions produced outputs that included extraneous markdown formatting (e.g., \texttt{```yaml} code blocks) or explanatory text alongside the YAML, requiring post-processing to extract the configuration. We addressed this by adding explicit constraints such as ``Output ONLY pure YAML content -- no explanations, markdown, or additional text.'' Early prompts also suffered from overly literal translations that preserved source CI syntax rather than adapting to target conventions; the instruction to ``avoid line-by-line literal translation in favor of capturing the pipeline's overall functionality'' was added to encourage semantic rather than syntactic migration. We also introduced the ``\texttt{\# TODO}'' comment requirement after observing that models would silently omit features lacking direct equivalents, making it difficult for users to identify incomplete migrations. These iterative refinements substantially improved output quality and usability across all evaluated models.

\begin{table}[ht]
\centering

\captionsetup{labelformat=empty}
\caption*{Prompt template used for {\approach}}
\vspace{-6pt}
\begin{minipage}{\linewidth}
\lstset{style=PromptStyle}

\begin{lstlisting}[label={lst:system-user-prompt-template},caption={Listing~\ref{lst:system-user-prompt-template}: Prompt template used for {\approach}},
  captionpos=t]
@u@System:@
@s@You are an Expert DevOps Engineer with extensive experience in GitHub Actions and Travis CI, with strong expertise in migrating CI/CD workflows from one service to another. Currently, you are tasked with migrating {{ SOURCE_CI }} to {{ TARGET_CI }} file.@

CORE REQUIREMENTS:
- Output ONLY pure YAML content - no explanations, markdown, or additional text
- Ensure strict {{ TARGET_CI }} syntax compliance
- Maintain semantic equivalence with the {{ SOURCE_CI }}
- Follow {{ TARGET_CI }} best practices and conventions
- Avoid line-by-line literal translation in favor of capturing the pipeline's overall functionality in the target workflow

@u@User:@
@s@Migrate this {{ SOURCE_CI }} to a {{ TARGET_CI }} YAML file.

REQUIREMENTS:
- Follow {{ TARGET_CI }} YAML syntax exactly
- Maintain all functionality from the original {{ SOURCE_CI }} configuration
- ...
- {Direction specific instructions / guidelines}
- ...
- Flag any {{ SOURCE_CI }} feature without a direct {{ TARGET_CI }} equivalent with a YAML comment (e.g., '# TODO: ...') for manual resolution
@
@d@{{ FEW_SHOT_EXAMPLES }}@
@s@
Now, do the migration and return ONLY the migrated valid {{ TARGET_CI }} YAML that accurately reflects the {{ SOURCE_CI }} shown below:
@
@d@{{ INPUT_CODE }}@
\end{lstlisting}

\end{minipage}

\vspace{-10pt}
\end{table}

\begin{table}[ht]
\centering
\captionsetup{labelformat=empty}
\caption*{Listing~\ref{lst:few_shot_gha_to_travis_example}: Few-shot example (input-output pair) used in prompting: Travis~CI$\rightarrow$GitHub~Actions}
\vspace{-6pt}
\begin{tabular}{C{5cm} | C{8cm}}
\multicolumn{1}{c}{\textsf{Travis~CI}} & \multicolumn{1}{c}{\textsf{GitHub~Actions}} \\
\hline
\lstset{style=yml}
\begin{lstlisting}[label={lst:few_shot_gha_to_travis_example},caption={~},aboveskip=-18pt,belowskip=-8pt]
|<deploy>|:
  |<provider>|: releases
  |<user>|: "GITHUB USERNAME"
  |<password>|: "GITHUB PASSWORD"
  |<file>|:
    - file-1
    - file-2
  |<draft>|: true
  |<overwrite>|: true
  |<tag_name>|: v2
  |<skip_cleanup>|: true
  |<prerelease>|: true
  |<name>|: Version dos
  |<body>|: This is the body
  |<on>|:
    |<tags>|: true
    
\end{lstlisting}
&
\lstset{style=yml}
\begin{lstlisting}[aboveskip=-8pt,belowskip=-8pt]
|<- uses>|: softprops/action-gh-release@v0.1.15
  |<env>|:
    |<GITHUB_TOKEN>|: "${{ secrets.GITHUB_TOKEN }}"
    |<GITHUB_REPOSITORY>|: "${{ github.repository }}"
  |<with>|:
    |<files>|: |-
      file-1
      file-2
    |<prerelease>|: true
    |<body>|: This is the body
    |<draft>|: true
    |<tag_name>|: v2
    |<name>|: Version dos
  |<if>|: "${{ github.event_name != 'pull_request' }}"
  
\end{lstlisting}

\\\\\hline
\end{tabular}
\vspace{-10pt}
\end{table}

\subsubsection{Zero-Shot Prompting}
\label{sec:zero_shot}

In zero-shot mode, we instantiate the prompt template in Listing~\ref{lst:system-user-prompt-template} by (i) setting \texttt{\{\{SOURCE\_CI\}\}} and \texttt{\{\{TARGET\_CI\}\}} according to the migration direction, (ii) injecting the direction-specific guideline block in the user message, and (iii) providing the source configuration via \texttt{\{\{INPUT\_CODE\}\}}. The placeholder \texttt{\{\{FEW\_SHOT\_EXAMPLES\}\}} is left empty. This setting evaluates whether an LLM can migrate CI configurations from instructions alone while meeting strict target-YAML constraints (format correctness) and preserving pipeline intent (semantic equivalence).

\subsubsection{Few-Shot Prompting}
\label{sec:few_shot}

In few-shot mode, we use the same system and user templates (Listing~\ref{lst:system-user-prompt-template}) but populate \texttt{\{\{FEW\_SHOT\_EXAMPLES\}\}} with three in-context migration pairs (3-shot), inserted immediately before \texttt{\{\{INPUT\_CODE\}\}}. The intent is to provide concrete demonstrations of how non-trivial CI constructs should be translated across services, which can improve structured generation through in-context learning~\cite{brown2020language}.

We select three in-context examples based on two criteria: (1) coverage of commonly occurring CI patterns, and (2) complementarity to avoid redundancy. While these examples do not exhaustively cover all possible CI configurations, they represent frequently used patterns that generalize well across diverse projects:
\begin{itemize}[leftmargin=*]
    \item \textbf{Example 1 (baseline-derived): Android build with caching.} A build-focused example illustrating environment setup and caching directives, sourced from migration artifacts used by the rule-based baseline approach~\cite{rzig2024cimig}.
    \item \textbf{Example 2 (external): GitHub release deployment.} A deployment example translating Travis CI releases (\texttt{deploy: provider: releases} with file attachments and metadata) into a GitHub Actions release step using a marketplace action, adapted from the GitHub Actions Importer documentation~\cite{actionsimporter_travis_release}.
    \item \textbf{Example 3 (external): Maven build with SonarCloud analysis.} A static-analysis example translating Travis CI \texttt{addons.sonarcloud} into explicit environment variables and a scan step in GitHub Actions, adapted from the GitHub Actions Importer documentation~\cite{actionsimporter_travis_sonarcloud}.
\end{itemize}

Together, these examples demonstrate environment configuration, caching strategies, deployment workflows, conditional execution, and third-party integrations. We deliberately avoided including multiple examples from the same domain (e.g., multiple Java builds) to maximize pattern diversity within the limited context window. This selection strategy follows established few-shot learning practices that prioritize complementary demonstrations over exhaustive coverage~\cite{brown2020language}.

For each migration direction, we provide direction-consistent input-output pairs (i.e., Travis CI$\rightarrow$GitHub Actions examples are used when migrating Travis CI inputs, and GitHub Actions$\rightarrow$Travis CI examples are used when migrating GitHub Actions inputs). Listing~\ref{lst:few_shot_gha_to_travis_example} corresponds to the \textbf{release-deployment} few-shot example (Example~2) used in our 3-shot prompt~\cite{actionsimporter_travis_release}. It illustrates the exact format in which a complete input-output migration pair is embedded inside the \texttt{\{\{FEW\_SHOT\_EXAMPLES\}\}} placeholder: a Travis CI \texttt{deploy: provider: releases} block with attached artifacts and release metadata is paired with its GitHub Actions counterpart that invokes \texttt{softprops/action-gh-release} and uses secrets-based authentication. This example is representative of the few-shot demonstrations we provide alongside the Android build-with-caching pair (Example~1) sourced from baseline migration artifacts~\cite{rzig2024cimig} and the Maven + SonarCloud integration pair (Example~3) adapted from the GitHub Actions Importer documentation~\cite{actionsimporter_travis_sonarcloud}. This setup tests whether demonstrations help models better preserve CI-specific idioms (e.g., triggers, secrets, and deployment semantics) while adhering to strict target syntax.

\subsubsection{Fine-Tuning}
\label{sec:fine_tuning}

To further improve migration quality and reduce reliance on in-context examples, we fine-tune \textbf{Gemma~3 (12B)} using supervised instruction tuning on a curated set of paired CI migrations. We keep the prompting interface consistent across training and testing by reusing the same system and user prompt structure shown in Listing~\ref{lst:system-user-prompt-template}. For each training instance, we instantiate \texttt{\{\{SOURCE\_CI\}\}} and \texttt{\{\{TARGET\_CI\}\}} according to the migration direction, inject the direction-specific guideline block in the user message, and provide the source configuration via \texttt{\{\{INPUT\_CODE\}\}}. During fine-tuning, the \texttt{\{\{FEW\_SHOT\_EXAMPLES\}\}} placeholder is not included in the prompt, so the model learns to produce target configurations directly from instructions and the input YAML.

Each training example is paired with a migrated configuration in the target CI format, which is used as the supervision target. This trains the model to satisfy the same constraints enforced at inference time: producing YAML-only outputs, adhering to strict target CI syntax, and preserving the pipeline's intended behavior.

\paragraph{Fine-tuning implementation details.}
We perform parameter-efficient fine-tuning using LoRA adapters~\cite{hu2022lora}, which update a small set of low-rank trainable parameters while keeping the base model weights frozen. This substantially reduces GPU memory requirements and training time, making the approach practical for iterative experimentation. We use the Unsloth framework~\cite{unsloth_finetuning,huggingface_unsloth_integration} rather than standard HuggingFace PEFT because Unsloth provides hand-written Triton kernels and a manual backpropagation engine that enable up to 2$\times$ faster training with up to 80\% less VRAM, while maintaining zero approximation error. Unsloth is fully compatible with HuggingFace's SFTTrainer, ensuring a streamlined workflow for LoRA-based instruction tuning. Table~\ref{tab:finetuning-hyperparams} summarizes the hyperparameters used for fine-tuning. The model is optimized with a next-token prediction objective on the target YAML conditioned on the concatenated system and user prompts. At inference time, we apply the same prompt template as in zero-shot prompting (again without \texttt{\{\{FEW\_SHOT\_EXAMPLES\}\}}), ensuring that any improvements reflect learned migration behavior rather than in-context demonstrations.

\begin{table}[t]
\centering
\caption{Hyperparameters used for fine-tuning Gemma~3 12B}
\label{tab:finetuning-hyperparams}
\begin{tabular}{lcc}
\toprule
\textbf{Hyperparameter} & \textbf{Travis$\rightarrow$GHA} & \textbf{GHA$\rightarrow$Travis} \\
\midrule
LoRA rank ($r$) & 64 & 64 \\
LoRA alpha ($\alpha$) & 16 & 16 \\
LoRA dropout & 0 & 0 \\
Learning rate & $10^{-4}$ & $10^{-3}$ \\
Batch size & 2 & 2 \\
Gradient accumulation steps & 4 & 4 \\
Epochs & 5 & 5 \\
Warmup ratio & 0.05 & 0.05 \\
Weight decay & 0.01 & 0.01 \\
Optimizer & AdamW & AdamW \\
\bottomrule
\end{tabular}
\end{table}

We conducted systematic hyperparameter search to identify optimal configurations for each migration direction, experimenting with learning rates in \{$10^{-5}$, $10^{-4}$, $10^{-3}$\}, varying batch sizes, and training epochs. Notably, the model required a 10$\times$ higher learning rate for the GHA$\rightarrow$Travis direction ($10^{-3}$ vs.\ $10^{-4}$) to achieve optimal results. This asymmetry reflects the inherent difficulty difference between the two tasks: mapping from GitHub Actions to Travis CI involves compressing a verbose, explicit configuration into a compact, implicit representation, which produces weaker or noisier gradient signals during training. The higher learning rate enables more decisive parameter updates within the limited training budget, whereas the Travis$\rightarrow$GHA expansion task converges effectively with a lower learning rate due to a smoother optimization landscape.

To avoid mixing incompatible YAML schemas within a single training run, we fine-tune using direction-consistent data (i.e., a run contains only Travis~CI$\rightarrow$GitHub~Actions pairs or only GitHub~Actions$\rightarrow$Travis~CI pairs). At test time, we evaluate using the same prompt template (without few-shot examples), enabling a direct comparison between prompting-only settings and fine-tuning under a matched prompting interface.

\subsection{Post-Processing and Validation}
\label{sec:post_processing}

Because migrated CI configurations are only actionable if they are syntactically valid and conform to the target service schema, {\approach} applies a lightweight two-stage validation pipeline after each generation.

\begin{itemize}[leftmargin=*]
    \item \textbf{Stage 1: YAML well-formedness (PyYAML).}
    We first parse every generated output using \texttt{PyYAML} to verify it is well-formed YAML~\cite{pyyaml}. This check is applied uniformly to both Travis CI and GitHub Actions outputs and filters out malformed generations early (e.g., indentation errors, invalid mappings, or broken multiline blocks).

    \item \textbf{Stage 2: Service-specific linting.}
    Only if YAML parsing succeeds, we run a service-aware linter to validate the configuration against platform-specific conventions and schema constraints. For GitHub Actions workflows, we use \texttt{actionlint}~\cite{actionlint} to detect issues such as invalid keys, missing required fields (e.g., \texttt{runs-on}), or malformed expressions. For Travis CI configurations, we use \texttt{travis-lint}~\cite{travis_lint} to validate \texttt{.travis.yml} structure and rule violations (e.g., unsupported sections or misformatted build matrices).
\end{itemize}

This staged design separates general YAML correctness from CI-specific validity, enabling us to report both basic parse success and stricter platform compliance in our evaluation.

\section{Empirical Evaluation}
\label{sec:evaluation}

\begin{figure*}
    \centering
    \vspace{-5pt}
    \includegraphics[width=1\textwidth]{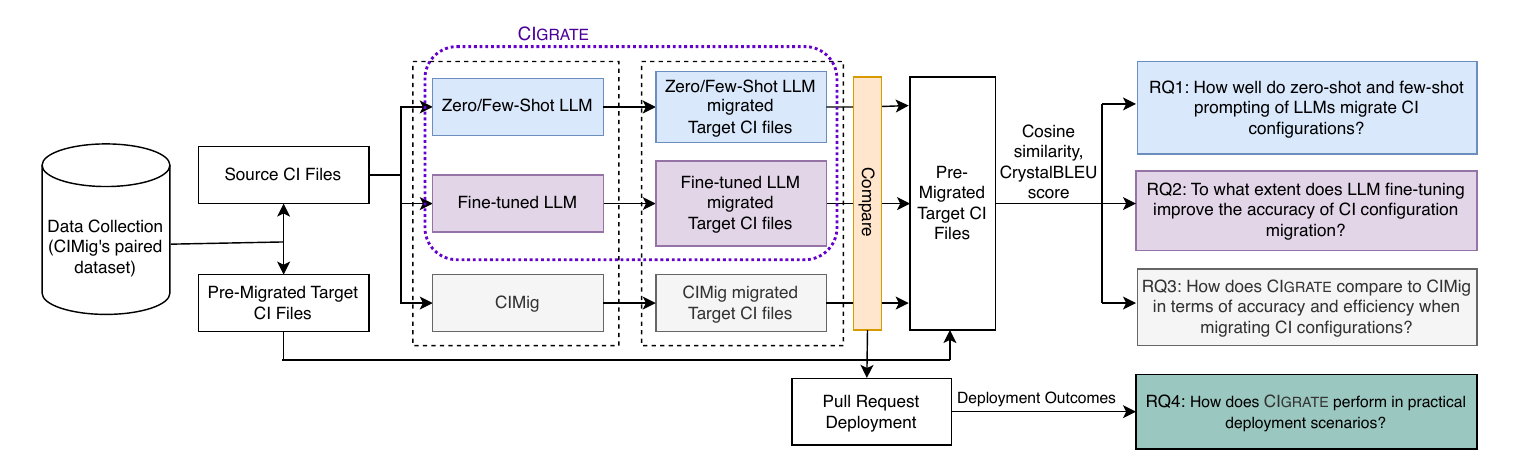} 
    \caption{Overview of our study}
    \vspace{-5pt}
    \label{fig:study_overview} 
\end{figure*}

This section reports on the experiments we conduct to assess the effectiveness of {\approach}. An overview of our study is presented in Figure~\ref{fig:study_overview}, summarizing the flow of our empirical evaluation used to address four research questions (RQs).

\subsection{Research Questions}

 \subsubsection{RQ1: How well do zero-shot and few-shot prompting of LLMs migrate CI configurations?}
 Zero-shot prompting offers a simple way to apply LLMs to CI configuration migration without requiring task-specific data. However, prior work has shown that performance can improve when LLMs are provided with in-context examples~\cite{brown2020language}. Few-shot prompting helps models better capture structural and formatting patterns relevant to configuration files. In this study, we use our {\approach} framework to systematically evaluate both zero-shot and few-shot prompting strategies and assess their effectiveness in automating CI configuration migration.

 \subsubsection{RQ2: To what extent does LLM fine-tuning improve the accuracy of CI configuration migration?}
 While zero-shot and few-shot LLMs offer practical and accessible solutions for CI configuration migration, their performance may be limited when handling complex, uncommon, or highly customized configurations. Fine-tuning injects domain-specific knowledge into the models, enabling them to better capture migration patterns, syntax nuances, and service-specific details. Prior studies have shown that task-specific fine-tuning can significantly enhance the accuracy and reliability of code-related tasks~\cite{wang2025llm,xu2023parameter}. 

 \subsubsection{RQ3: How does {\approach} compare to {\prevapproach} in terms of accuracy and effectiveness when migrating CI configurations?}
 {\prevapproach}~\cite{rzig2024cimig} represents the state-of-the-art in automating CI configuration migration through rule-based and example-driven techniques. However, its reliance on predefined mappings and extracted examples may limit its ability to generalize to unseen or more complex configurations. In contrast, LLM-based solutions like {\approach} offer a more flexible alternative by leveraging large-scale pretraining and contextual understanding. This RQ compares the performance of {\approach} and {\prevapproach} in terms of accuracy, feature preservation, and practical usability. Such a comparison helps determine whether LLMs can effectively complement or even replace rule-based solutions in this domain.

 \subsubsection{RQ4: How does {\approach} perform in practical deployment scenarios?}
 While automated metrics provide valuable insights, the ultimate test of a CI migration tool is whether its outputs work in real-world deployments. This RQ evaluates {\approach}'s practical usability by deploying generated configurations to active open-source projects, measuring CI execution success rates, and soliciting developer feedback through pull requests. Following prior work such as {\prevapproach}~\cite{rzig2024cimig}, this evaluation focuses on the generated outputs rather than the migration process itself.

\subsection{Baseline approach (\prevapproach)}

We compare {\approach} against {\prevapproach}~\cite{rzig2024cimig}, the state-of-the-art rule-based approach for CI configuration migration. {\prevapproach} operates by mining transformation rules from a large corpus of aligned CI configuration pairs. It uses abstract syntax tree (AST) matching and association rule mining to learn mappings between Travis CI and GitHub Actions constructs. When given a source configuration, {\prevapproach} applies the learned rules to generate an equivalent target configuration.

While {\prevapproach} demonstrated the feasibility of automated CI migration, it has several limitations that motivate our LLM-based approach. First, its rule coverage depends on the diversity of training examples; configurations with uncommon features or complex job matrices may not have corresponding rules. Second, it struggles with semantic preservation when features lack direct structural equivalents (e.g., certain GitHub Actions triggers or Travis addons). Third, as we demonstrate in our evaluation, {\prevapproach} produces syntactically invalid YAML in the vast majority of cases, severely limiting its practical utility.

\subsection{Dataset}
\label{sec:dataset}

Our evaluation uses a curated dataset derived from the {\prevapproach} corpus~\cite{rzig2024cimig}. {\prevapproach} collected 1,252 dual-CI projects (using both Travis CI and GitHub Actions) after applying filtering based on git-history analysis, cosine similarity thresholds, and manual verification of semantic equivalence, then reserved 251 projects as the evaluation set following an 80/20 train-test split. We adopt this 251-project evaluation set as our starting point. However, upon careful examination, we identified significant quality issues in the original dataset. Many pairs exhibited poor functional equivalence because the GitHub Actions workflow and Travis CI configuration were captured at different points in the project's evolution, sometimes years apart, resulting in substantial semantic drift between the paired files.

To address this limitation, we constructed a high-quality ground-truth dataset through the following process. First, we computed the minimum sample size required for statistical validity using a sample size calculator with a 95\% confidence level, $\pm$5\% margin of error, and a population of 251 projects, yielding a required sample of 153 projects. We then randomly selected 153 projects from the original 251 and enlisted two software engineering researchers with CI/CD expertise to collaboratively create functionally equivalent GitHub Actions workflows for each Travis CI configuration through joint review meetings. This manual curation ensures that our ground-truth pairs represent semantically faithful migrations rather than historically coincidental pairings.

For the zero-shot and few-shot experiments, we evaluated all 153 projects. For fine-tuning, we adopted a 10-fold cross-validation protocol to obtain a robust estimate of generalization performance on unseen projects while maximizing the use of the limited curated dataset. Concretely, we partitioned the 153 projects into 10 disjoint folds and repeated training/testing ten times, each time holding out one fold as the test set and using the remaining folds for training. We use a 90/10 split per run, corresponding to 15 projects in the test fold and 138 projects for training (with the remaining 0 to 1 project handled by balancing to keep fold sizes as even as possible). Metrics are computed on the held-out fold and then averaged across the 10 runs, which reduces sensitivity to a single train/test partition and provides variance estimates across folds. In total, this protocol yields 150 held-out evaluations (15 projects $\times$ 10 folds), with all curated projects participating in testing exactly once across the cross-validation process.

Our evaluation is based on the manually curated dataset rather than the full set of 1,252 workflow pairs because the original corpus contained substantial semantic drift between many paired Travis CI and GitHub Actions files. Evaluating against these stale references would reward migrations for matching outdated workflows rather than producing semantically correct translations. After reconstructing a high-quality ground truth at the statistically justified sample size of 153 projects, 10-fold cross-validation provided a more robust estimate of generalization than a single fixed train/test split while making efficient use of the limited curated dataset.

\subsection{Practical Deployment Evaluation}
\label{sec:pull_requests}
While automated metrics provide valuable insights into migration accuracy, the ultimate measure of a migration tool's utility is whether developers find its outputs acceptable for real-world use. To assess this, we conducted a developer feedback study by submitting pull requests (PR) containing {\approach}-generated CI configurations to active open-source projects on GitHub.

We selected 30 GitHub projects for the pull request study, divided equally between the two migration directions (15 projects each). For each direction, we applied strict selection criteria to ensure meaningful evaluation:

\paragraph{GHA$\rightarrow$Travis Migration (15 projects).} We identified projects that have used GitHub Actions throughout their entire lifetime and have never used Travis CI. For these projects, we used the {\approach} fine-tuned model to generate an equivalent Travis CI configuration from their existing GitHub Actions workflow. This scenario represents a realistic use case where a project might want to add Travis CI support (e.g., for testing on different infrastructure or as a backup CI service).

\paragraph{Travis$\rightarrow$GHA Migration (15 projects).} Conversely, we selected projects that have exclusively used Travis CI and have never adopted GitHub Actions. We generated GitHub Actions workflows using the {\approach} fine-tuned model, representing the common scenario of projects modernizing their CI infrastructure by migrating to GitHub Actions.

\paragraph{PR validation and feedback collection.}
Before submitting PRs, we validated each generated configuration using a two-stage check: YAML parsing and service-specific linting (i.e., \texttt{actionlint}\cite{actionlint} for GitHub Actions workflows and \texttt{travis-lint}\cite{travis_lint} for Travis CI configurations). We also tried to run the generated workflows/configurations on their respective platforms (GitHub Actions and Travis CI) to ensure they could be parsed and initiated by the CI service.

Each PR included the generated configuration file and requested maintainers to provide a five-point rating of the migration quality with respect to the source CI configuration, along with qualitative comments (e.g., missing steps, incorrect triggers, or project-specific constraints). For PRs where the generated configuration passed CI execution, we highlighted this success; for unsuccessful runs, we acknowledged the issue while still requesting structural feedback. To maximize response rate, we followed up multiple times after submission by posting reminder comments on the PR threads. We also explicitly welcomed feedback from developers who had no immediate plans to migrate but were willing to review the generated configuration for quality and maintainability.

We originally aimed to evaluate {\approach} through maintainers' five-point ratings, analyzed using the Wilcoxon signed-rank test, with CI execution outcomes serving as a complementary signal. As reported in Section~\ref{sec:rq4-results}, maintainer response to unsolicited PRs was limited (19 of 30 PRs received no response, and only one PR included an explicit numeric rating), preventing a meaningful Likert-scale statistical analysis. We therefore report CI execution success as the primary quantitative outcome for RQ4, while maintainer ratings and qualitative feedback, where available, are presented descriptively rather than subjected to statistical testing. The implications of this change are discussed in Section~\ref{sec:threats_to_validity}.
    
\subsection{Evaluation metrics}

We evaluate {\approach} using two categories of metrics: similarity metrics that measure semantic fidelity to the ground truth, and linting metrics that assess syntactic validity.

\subsubsection{Similarity Metrics}

We employ two complementary similarity metrics, following established practice in code generation evaluation~\cite{eghbali2022crystalbleu}:

\paragraph{Cosine Similarity.} Cosine Similarity measures the angular distance between two text vectors in a high-dimensional space~\cite{gunawan2018cosine}. We tokenize both the generated and reference configurations, convert them to TF-IDF vectors, and compute the cosine of the angle between them. This metric captures overall structural similarity, with values ranging from 0 (completely different) to 1 (identical).

\paragraph{CrystalBLEU.} CrystalBLEU is a variant of the BLEU score specifically designed for code evaluation~\cite{eghbali2022crystalbleu}. Unlike standard BLEU, CrystalBLEU down-weights trivially shared tokens (e.g., common keywords, punctuation) that would otherwise inflate similarity scores. This metric better captures meaningful semantic similarity in configuration files, where boilerplate tokens are common.

\subsubsection{Linting Metrics}

Beyond semantic similarity, we assess whether generated configurations are syntactically valid and ready for practical use:

\paragraph{YAML Validity.} We parse each generated configuration using a standard YAML parser to verify basic structural correctness. Invalid YAML (e.g., malformed indentation, invalid characters) cannot be used in any CI service and represents a fundamental failure of the migration tool.

\paragraph{CI Linter Pass Rate.} We apply service-specific linters to validate semantic correctness against each platform's schema and conventions. For GitHub Actions, we use \texttt{actionlint}, which checks workflow structure, action references, expression syntax, and other GHA-specific requirements. For Travis CI, we use \texttt{travis-lint}, which validates configuration against the Travis CI schema. These linters catch issues that would cause CI failures even if the YAML is technically parseable.

\subsection{Statistical Testing}

To assess the statistical significance of performance differences, we use the Wilcoxon signed-rank test~\cite{woolson2007wilcoxon}, a non-parametric test appropriate for paired samples that does not assume normal distribution of differences. We perform one-sided tests (alternative = ``greater'') to determine whether {\approach} significantly outperforms {\prevapproach} on each metric. We use a significance threshold of $\alpha = 0.05$ and report exact p-values for transparency.

\section{Results}
\label{sec:results}

This section presents the results of our empirical evaluation, organized by research question. For each RQ, we report respective quantitative metrics (such as Cosine Similarity, CrystalBLEU, and linting pass rates) along with statistical significance tests. For brevity, we sometimes report paired similarity values in-text using the shorthand \textit{Cosine/CrystalBLEU} (e.g., 0.66/0.44); otherwise, we report the two metrics explicitly. All pairwise comparisons against the CIMig baseline use the Wilcoxon signed-rank test with a significance threshold of $\alpha = 0.05$.

\subsection{RQ1 Results}
\label{sec:rq1-results}

\begin{figure}[t]
\centering
\includegraphics[width=.89\textwidth]{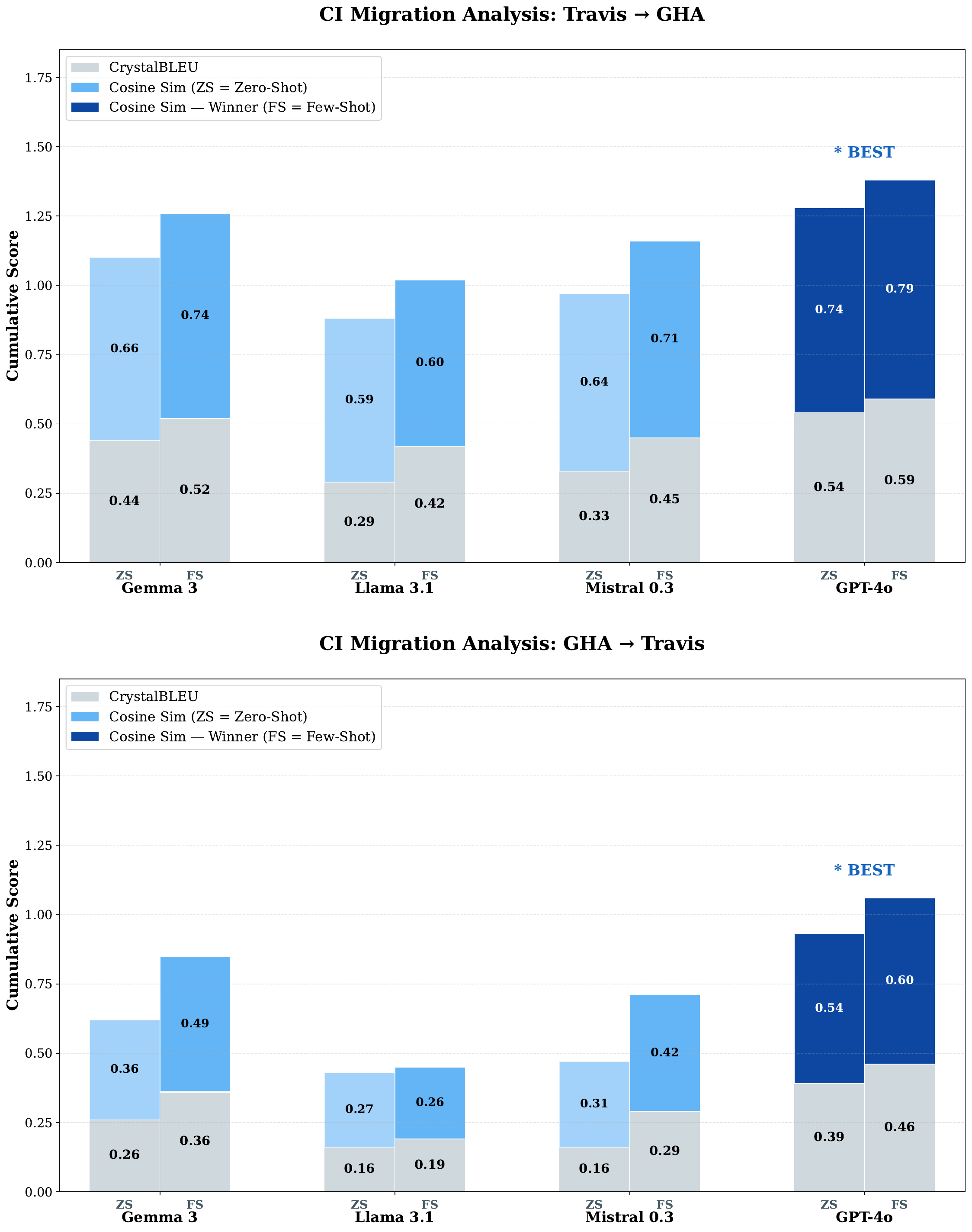}
\caption{Comparison of Cosine similarity and CrystalBLEU across zero-shot and few-shot prompting for the four evaluated LLMs, reported separately for Travis$\rightarrow$GHA and GHA$\rightarrow$Travis.}
\label{fig:rq1_similarity_comparison}
\end{figure}

We evaluate how well \emph{zero-shot} and \emph{few-shot} prompting support CI configuration migration. We consider four LLMs (GPT-4o, Gemma~3 12B, Mistral 7B, and Llama~3.1 8B) and perform bidirectional migration between Travis~CI and GitHub~Actions over 153 migration pairs. Table~\ref{tab:rq1-performance} reports the mean Cosine Similarity and CrystalBLEU scores for each model and prompting setting. Figure~\ref{fig:rq1_similarity_comparison} visualizes the same comparison across models and directions.

\subsubsection{Travis CI $\rightarrow$ GitHub Actions}

For Travis$\rightarrow$GHA migration, GPT-4o achieves the highest similarity scores across both metrics and benefits from few-shot prompting, improving from 0.74/0.54 (Cosine/CrystalBLEU) to 0.79/0.59. Among the open-weight models, few-shot prompting consistently improves Gemma from 0.66/0.44 to 0.74/0.52 and Mistral from 0.64/0.33 to 0.71/0.45. Llama 3.1 exhibits only a marginal improvement in Cosine Similarity (0.59 to 0.60) but a larger gain in CrystalBLEU (0.29 to 0.42), indicating that few-shot prompting benefits this model more under CrystalBLEU than under Cosine Similarity. Statistical analysis using the Wilcoxon signed-rank test shows that all few-shot improvements are significant ($p < 0.001$, Bonferroni-corrected $\alpha = 0.0031$), with medium to large matched-pairs rank-biserial effect sizes ($r = 0.43$ to $0.76$), except for Llama 3.1's Cosine Similarity, which exhibits a negligible effect size ($r = 0.03$).

\subsubsection{GitHub Actions $\rightarrow$ Travis CI}

GitHub Actions$\rightarrow$Travis CI migration is more challenging overall, reflecting the need to translate GitHub Actions' explicit workflow structure into Travis CI's more compact phase-based format. GPT-4o again achieves the highest similarity scores across both metrics and benefits from few-shot prompting, improving from 0.54/0.39 (Cosine/CrystalBLEU) to 0.60/0.46. Among the open-weight models, few-shot prompting consistently improves Gemma from 0.36/0.26 to 0.49/0.36 and Mistral from 0.31/0.16 to 0.42/0.29. In contrast, Llama 3.1 exhibits limited improvement in this direction, with Cosine Similarity decreasing slightly from 0.27 to 0.26 while CrystalBLEU increases modestly from 0.16 to 0.19. These results suggest that GitHub Actions$\rightarrow$Travis CI migration is more difficult for smaller models, likely because it requires condensing GitHub Actions' richer workflow representation into Travis CI's simpler syntax. Statistical analysis using the Wilcoxon signed-rank test shows that most few-shot improvements are significant ($p < 0.001$, Bonferroni-corrected $\alpha = 0.0031$), with medium to large matched-pairs rank-biserial effect sizes ($r = 0.33$ to $0.71$). The only exception is Llama 3.1's Cosine Similarity, which exhibits a negligible negative effect size ($r = -0.01$).

\subsubsection{Syntactic Validity}

Beyond similarity, we assess whether the generated configurations are syntactically usable using (i) YAML parsing and (ii) CI-specific linters (actionlint for GitHub Actions outputs and travis-lint for Travis outputs). Overall, prompting with GPT-4o yields high syntactic validity (about 99\% YAML validity and strong linter pass rates in both directions), while Gemma and Mistral also produce valid outputs in many cases. One notable exception is Llama~3.1 8B, where few-shot prompting can reduce syntactic correctness: YAML validity drops from 28.1\% (zero-shot) to 2.0\% (few-shot) for Travis$\rightarrow$GHA, and from 24.2\% to 0.7\% for GHA$\rightarrow$Travis.

\paragraph{Failure-mode analysis for Llama~3.1 8B.}
We classified all 612 Llama outputs (306 few-shot, 306 zero-shot; 153 per direction) with an automated open-coding script (\texttt{analyze\_llama\_failures.py}) and spot-checked representative cases. Among 302 invalid few-shot outputs, 96\% (290/302) contained markdown code fences and conversational preamble (e.g., \textit{``Here is the migrated GitHub Actions workflow:''} followed by \texttt{```yaml}), directly mirroring the markdown-wrapped structure of our few-shot examples rather than following the instruction to emit pure YAML. A secondary failure mode (18\% of Travis$\rightarrow$GHA few-shot failures) copied idioms from the in-context demonstrations---e.g., \texttt{jdk: ['8', '9']} and \texttt{distribution: temurin}---even when absent from the source Travis file. For GHA$\rightarrow$Travis few-shot, 13\% of failures partially echoed the input GitHub Actions workflow instead of condensing it into Travis syntax. Zero-shot failures were more heterogeneous: duplicate YAML keys (38--47\%), markdown fences without preamble (30--36\%), and unmigrated input fragments (18\%). Few-shot prompting therefore degrades Llama primarily through \emph{format non-compliance}, not through lower migration fidelity---CrystalBLEU still improves significantly for Llama in both directions ($p < 0.001$), while Cosine Similarity gains are negligible ($p > 0.12$, $r \leq 0.03$). Input length showed only a weak association with failure (median 20 vs.\ 12 lines for failed vs.\ successful Travis$\rightarrow$GHA few-shot outputs), suggesting that context-window pressure is not the dominant factor.

\begin{table}[t]
\centering
\caption{RQ1 performance: Mean Cosine Similarity and CrystalBLEU scores for {\approach} under zero-shot and few-shot prompting across 153 projects. Results are reported separately for each migration direction. Boldface indicates the best result per metric and direction.}
\label{tab:rq1-performance}
\begin{tabular}{llcccc}
\toprule
\textbf{Model} & \textbf{Prompt} &
\multicolumn{2}{c}{\textbf{Travis $\rightarrow$ GHA}} &
\multicolumn{2}{c}{\textbf{GHA $\rightarrow$ Travis}} \\
\cmidrule(lr){3-4} \cmidrule(lr){5-6}
& & \textit{Cosine} & \textit{CrystalBLEU}
& \textit{Cosine} & \textit{CrystalBLEU} \\
\midrule
GPT-4o & Zero-shot
& 0.74 & 0.54
& 0.54 & 0.39 \\
 & Few-shot
& \textbf{0.79} & \textbf{0.59}
& \textbf{0.60} & \textbf{0.46} \\

\midrule
Gemma 3 12B & Zero-shot
& 0.66 & 0.44
& 0.37 & 0.26 \\
 & Few-shot
& 0.74 & 0.52
& 0.49 & 0.36 \\

\midrule
Mistral 7B & Zero-shot
& 0.64 & 0.33
& 0.31 & 0.16 \\
 & Few-shot
& 0.71 & 0.45
& 0.42 & 0.29 \\

\midrule
Llama 3.1 8B & Zero-shot
& 0.59 & 0.29
& 0.27 & 0.16 \\
 & Few-shot
& 0.60 & 0.42
& 0.26 & 0.19 \\
\bottomrule
\end{tabular}
\vspace{-7pt}
\end{table}

\summarybox{\textbf{RQ1 Summary.}
Zero-shot prompting already enables meaningful CI migration across all models, while few-shot prompting generally improves similarity to reference configurations. GPT-4o performs best overall and improves with few-shot prompting in both directions (Travis$\rightarrow$GHA: 0.74/0.54 $\rightarrow$ 0.79/0.59; GHA$\rightarrow$Travis: 0.54/0.39 $\rightarrow$ 0.60/0.46 in Cosine/CrystalBLEU). Open-weight models (Gemma, Mistral) also benefit from few-shot prompting, indicating that a small number of examples helps capture CI conventions. Across all models, GHA$\rightarrow$Travis is consistently harder than Travis$\rightarrow$GHA. However, few-shot prompting is not universally beneficial, sometimes reducing stability in smaller models (e.g., Llama), suggesting it should be used more cautiously for low-capacity models.
}

\subsection{RQ2 Results}
\label{sec:rq2-results}

Building on the RQ1 findings, we assess whether domain-specific fine-tuning improves CI migration accuracy beyond prompting-only setups. We fine-tune Gemma~3 12B (the best-performing open-weight model in RQ1) on the manually curated ground-truth dataset. We employ 10-fold cross-validation with a 90/10 train/test split to obtain a robust estimate of generalization performance, yielding 150 test samples in total (15 per fold). We compare the fine-tuned model against the strongest prompting-only baseline from RQ1, namely GPT-4o (zero-shot and few-shot). Table~\ref{tab:rq2-finetuning} summarizes the comparison, and Figure~\ref{fig:finetuning-folds-boxplot} visualizes the fold-level distribution of similarity scores for the fine-tuned model.

\subsubsection{Travis CI $\rightarrow$ GitHub Actions}

Fine-tuning clearly improves performance in the Travis$\rightarrow$GHA direction: the fine-tuned Gemma~3 12B achieves a mean Cosine Similarity of \textbf{0.90} ($\pm$0.02) and a CrystalBLEU of \textbf{0.74} ($\pm$0.04). Compared to the strongest prompting-only baseline (GPT-4o few-shot, 0.79/0.59), fine-tuning provides an absolute gain of +0.11 in Cosine and +0.15 in CrystalBLEU. Relative to GPT-4o zero-shot (0.74/0.54), the improvement is larger (+0.16 Cosine, +0.20 CrystalBLEU). Mann-Whitney U tests confirm these improvements are statistically significant ($p < 0.001$, Bonferroni-corrected $\alpha = 0.0042$), with large effect sizes ($\delta = 0.90$ for Cosine, $\delta = 0.74$ for CrystalBLEU). The compact spread of the Travis$\rightarrow$GHA box plots in Figure~\ref{fig:finetuning-folds-boxplot} indicates that these gains are stable across folds rather than due to a single favorable split.

\subsubsection{GitHub Actions $\rightarrow$ Travis CI}

Fine-tuning also improves the more challenging GHA$\rightarrow$Travis direction. The fine-tuned Gemma~3 12B achieves \textbf{0.76} Cosine ($\pm$0.06) and \textbf{0.65} CrystalBLEU ($\pm$0.06). Compared to GPT-4o few-shot (0.60/0.46), fine-tuning yields +0.16 Cosine and +0.19 CrystalBLEU. These improvements are also statistically significant ($p < 0.001$, Bonferroni-corrected $\alpha = 0.0042$), with large effect sizes ($\delta = 0.82$ for Cosine, $\delta = 0.80$ for CrystalBLEU). Figure~\ref{fig:finetuning-folds-boxplot} shows slightly higher fold-to-fold variability than in Travis$\rightarrow$GHA, consistent with the additional difficulty of translating GitHub Actions' more explicit workflow structure into Travis~CI's compact phase-based format. We also observed that this direction required a higher learning rate during fine-tuning ($10^{-3}$ vs.\ $10^{-4}$ for Travis$\rightarrow$GHA) to reach the best-performing configuration.

\begin{table}[t]
\centering
\caption{RQ2 performance: Fine-tuned Gemma~3 12B results (10-fold cross-validation, $n=150$) compared to the best prompting-only baselines (GPT-4o, $n=153$). Standard deviations for fine-tuning reflect variation across folds.}
\label{tab:rq2-finetuning}
\begin{tabular}{llcccc}
\toprule
\textbf{Model} & \textbf{Setting} &
\multicolumn{2}{c}{\textbf{Travis $\rightarrow$ GHA}} &
\multicolumn{2}{c}{\textbf{GHA $\rightarrow$ Travis}} \\
\cmidrule(lr){3-4} \cmidrule(lr){5-6}
& & \textit{Cosine} & \textit{CrystalBLEU}
& \textit{Cosine} & \textit{CrystalBLEU} \\
\midrule
GPT-4o & Zero-shot
& 0.74 & 0.54
& 0.54 & 0.39 \\
 & Few-shot
& 0.79 & 0.59
& 0.60 & 0.46 \\
\midrule
Gemma 3 12B & Fine-tuned
& \textbf{0.90}$\pm$0.02 & \textbf{0.74}$\pm$0.04
& \textbf{0.76}$\pm$0.06 & \textbf{0.65}$\pm$0.06 \\
\bottomrule
\end{tabular}
\vspace{-7pt}
\end{table}

\subsubsection{Fold-level Distribution}

\begin{figure}[t]
    \centering
    \includegraphics[width=.89\textwidth]{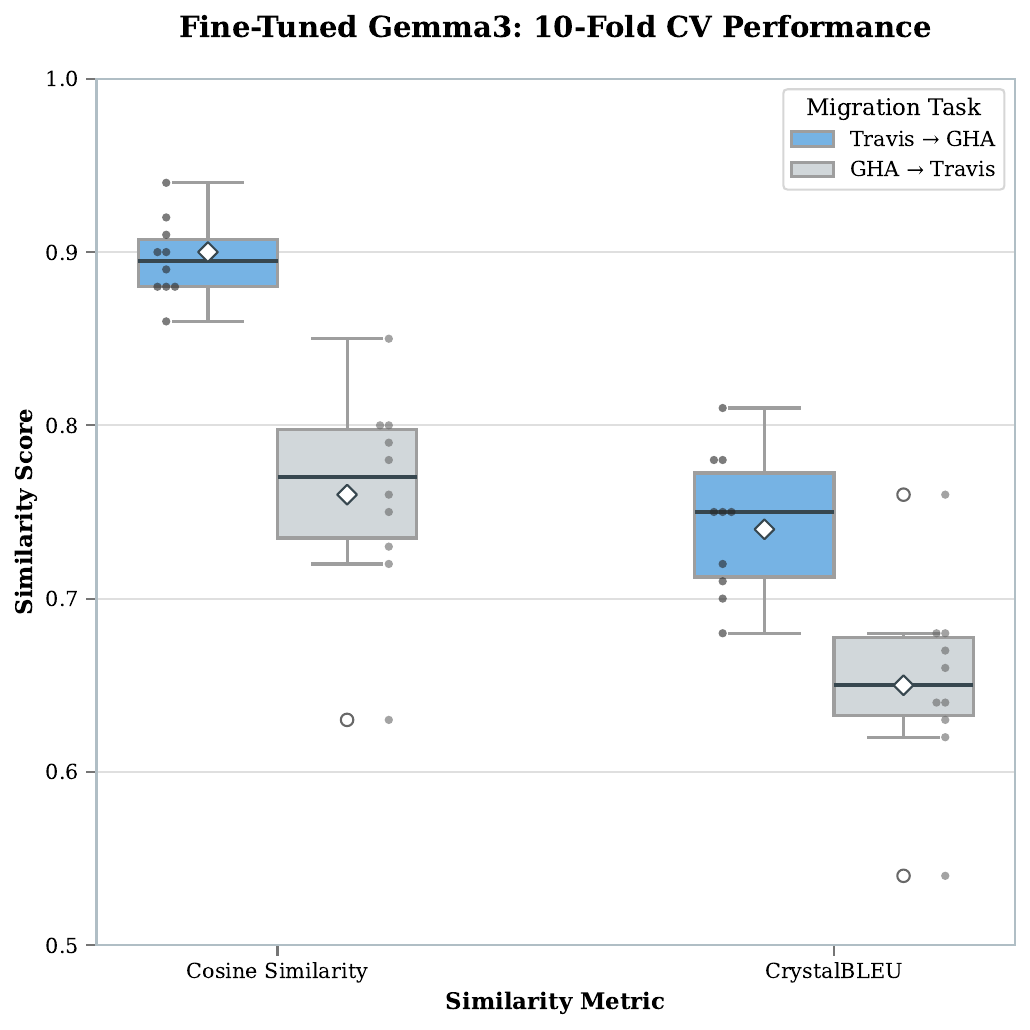}
    \caption{Performance of the fine-tuned Gemma~3 12B model across 10-fold cross-validation. Box plots show the distribution of similarity scores across folds for Cosine Similarity and CrystalBLEU under both migration directions. White diamond markers denote the mean, and individual points represent fold-level results.}
    \label{fig:finetuning-folds-boxplot}
\end{figure}

Figure~\ref{fig:finetuning-folds-boxplot} summarizes the distribution of fine-tuning results across the 10 folds for each direction and metric. The plots show the median and interquartile range across folds, with mean markers, making robustness and dispersion visually apparent. Overall, Travis$\rightarrow$GHA exhibits a tighter distribution for both metrics (per-project cosine $\sigma = 0.09$; fold-mean $\sigma = 0.02$), while GHA$\rightarrow$Travis shows a wider spread (per-project cosine $\sigma = 0.20$; fold-mean $\sigma = 0.06$; fold means ranging from 0.63 to 0.85), aligning with its higher standard deviation in Table~\ref{tab:rq2-finetuning}. The weakest fold (\texttt{Set\_90\_10\_V\_10}, mean cosine 0.63) was driven by projects such as \texttt{JodaOrg\_joda-time} (0.28), \texttt{constellation-app\_constellation} (0.34), \texttt{OpenLiberty\_ci.maven} (0.36, build matrix), and \texttt{vert-x3\_vertx-bridge-common} (0.36, multi-job workflow), indicating that compression difficulty concentrates in workflows with matrices and multiple jobs rather than in a single anomalous project. Comparing the bottom vs.\ top 20\% of GHA$\rightarrow$Travis cosine scores ($n=30$ each), low-scoring source workflows were longer (45 vs.\ 31 lines), used more steps (1.6 vs.\ 1.1), and more often contained build matrices (23\% vs.\ 10\%), conditional \texttt{if:} guards (1.4 vs.\ 0.2 per workflow), \texttt{secrets.*} references (0.7 vs.\ 0.03), and \texttt{GITHUB\_TOKEN} usage (13\% vs.\ 3\%). Top-level \texttt{permissions:} blocks were rare in our corpus ($<$2\% of workflows) and did not differentiate the groups.

\subsubsection{Syntactic Validity Under Fine-Tuning}

Fine-tuning improves not only similarity but also output robustness. The fine-tuned model achieves 100\% YAML validity for Travis$\rightarrow$GHA outputs and 97.3\% for GHA$\rightarrow$Travis, with corresponding lint pass rates of 89.3\% and 97.3\%. These results indicate that fine-tuning helps internalize target schema conventions and reduces malformed outputs, making the generated configurations more likely to run with minimal manual fixes.

\summarybox{\textbf{RQ2 Summary.}
Fine-tuning consistently outperforms the strongest prompting-only baseline. Compared to GPT-4o few-shot, the fine-tuned Gemma~3 12B improves similarity from 0.79/0.59 to \textbf{0.90/0.74} for Travis$\rightarrow$GHA and from 0.60/0.46 to \textbf{0.76/0.65} for GHA$\rightarrow$Travis (Cosine/CrystalBLEU). Gains are larger for the harder GHA$\rightarrow$Travis direction, indicating that fine-tuning better captures direction-specific restructuring patterns than prompting alone. Across 10 folds, results show stable generalization with low variance for Travis$\rightarrow$GHA and slightly higher variance for GHA$\rightarrow$Travis. Overall, prompting provides strong baseline performance, while fine-tuning yields the most accurate and reliable migrations when representative training data is available.
}

\subsection{RQ3 Results}
\label{sec:rq3-results}
\begin{figure}[t]
        \centering
        \includegraphics[width=.9\textwidth]{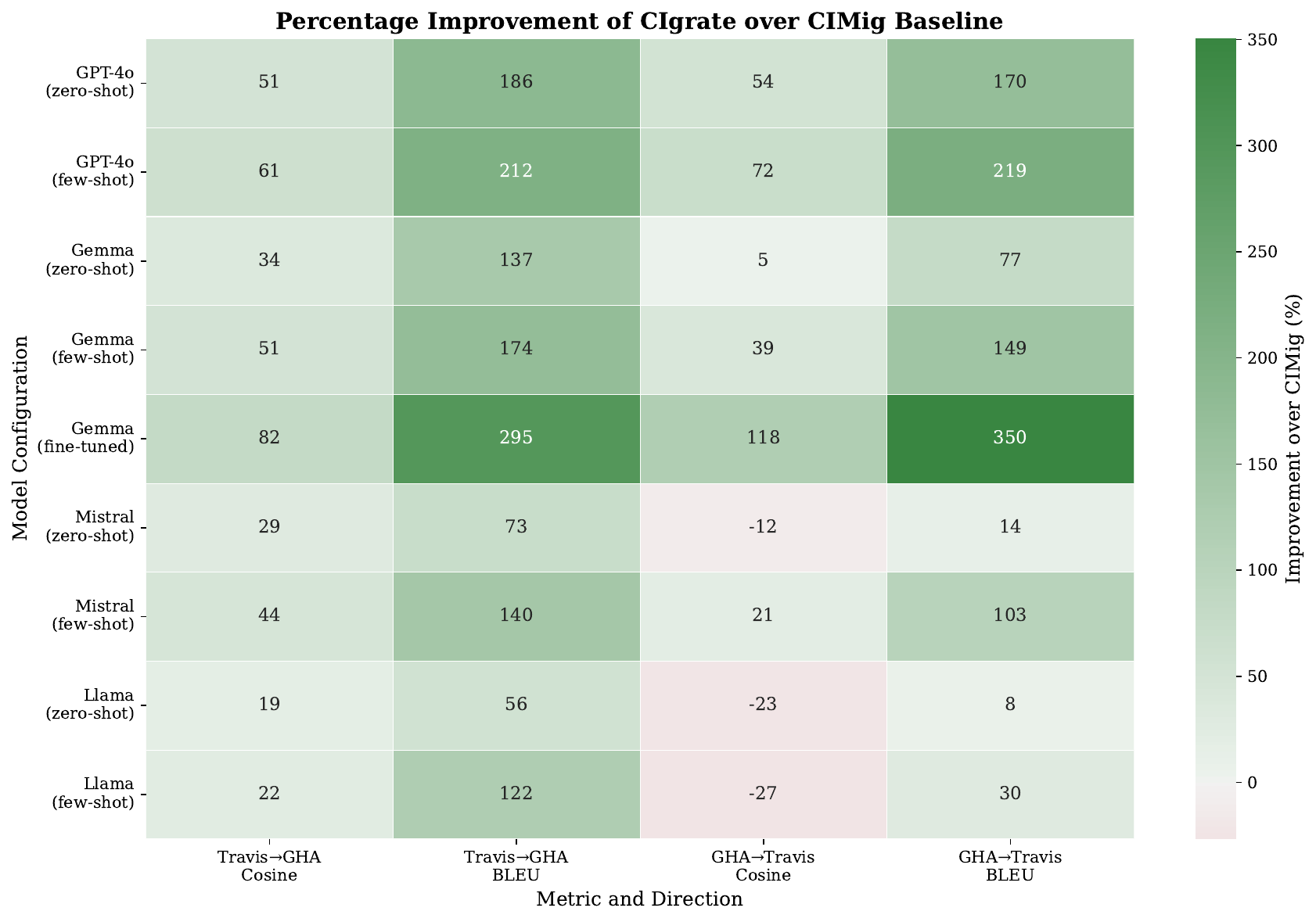}
        \caption{Percentage improvement of {\approach} configurations over the {\prevapproach} baseline. Green indicates improvement; red indicates underperformance. The fine-tuned Gemma model achieves the largest improvements (+82\% to +350\%), over \prevapproach.}
        \label{fig:improvement-heatmap}
    \end{figure}

    This RQ directly compares {\approach} against the state-of-the-art rule-based baseline, {\prevapproach}, across all configurations. Table~\ref{tab:rq3-comparison} consolidates the comparison, and Figure~\ref{fig:improvement-heatmap} visualizes the percentage improvements.

    \subsubsection{Similarity Metrics}
    
    Across both directions, every model significantly outperforms {\prevapproach} on at least one metric, with the Travis$\rightarrow$GHA direction yielding uniform improvements and GHA$\rightarrow$Travis showing more model- and prompt-dependent results.

    \textbf{Travis$\rightarrow$GHA.} Against {\prevapproach}'s baseline of 0.49/0.19 (Cosine/CrystalBLEU), every {\approach} configuration, zero-shot, few-shot, and fine-tuned, achieves statistically significant improvements on both metrics (all $p < 0.001$, Wilcoxon signed-rank test). Even the weakest configuration, LLaMA~3.1 8B zero-shot, reaches 0.59/0.29 ($p < 10^{-11}$/$p < 10^{-7}$, $r = 0.57$/$r = 0.33$, large/medium effects), while the strongest prompting-only configuration, GPT-4o few-shot, achieves 0.79/0.59 ($p < 0.001$, $r = 0.99$/$r = 0.97$, large effects).

    \textbf{GHA$\rightarrow$Travis.} This direction is more challenging overall, with results that vary substantially by model and prompt mode. Against {\prevapproach}'s baseline of 0.35/0.14 (Cosine/CrystalBLEU), GPT-4o consistently outperforms on both metrics regardless of prompt mode: zero-shot reaches 0.54/0.39 ($p < 10^{-19}$/$p < 10^{-23}$, $r = 0.70$/$r = 0.80$, large effects) and few-shot extends this to 0.60/0.46 ($p < 10^{-23}$/$p < 10^{-25}$, $r = 0.80$/$r = 0.90$, large effects). Gemma~3 12B and Mistral 7B improve significantly only with few-shot prompting: Gemma few-shot reaches 0.49/0.36 ($p < 10^{-12}$/$p < 10^{-22}$, $r = 0.48$/$r = 0.81$, medium/large effects) and Mistral few-shot 0.42/0.29 ($p < 10^{-5}$/$p < 10^{-16}$, $r = 0.31$/$r = 0.62$, medium/large effects). In zero-shot mode, Gemma improves only CrystalBLEU significantly ($p < 10^{-10}$, $r = 0.43$, medium effect) while its Cosine score of 0.37 does not significantly exceed the baseline; Mistral zero-shot improves neither. LLaMA~3.1 8B fails to improve on either metric in zero-shot mode and only marginally improves CrystalBLEU in few-shot mode ($p < 10^{-3}$, $r = 0.21$, small effect), with Cosine scores (0.27/0.26) falling \emph{below} the {\prevapproach} baseline, indicating this direction is particularly challenging for smaller models.
    
\begin{figure}[t]
        \centering
        \includegraphics[width=.89\textwidth]{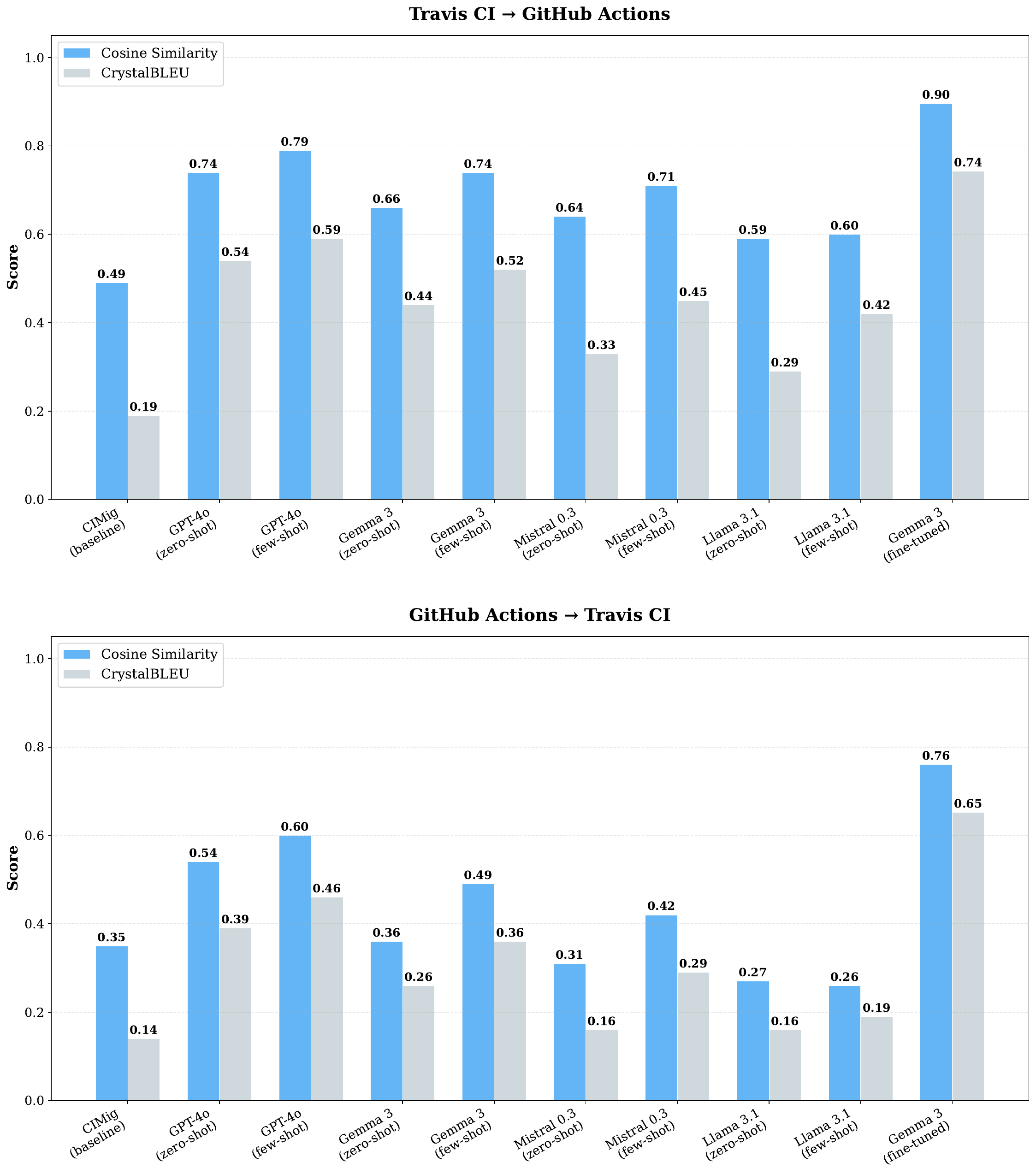}
        \caption{Comparison of Cosine Similarity and CrystalBLEU scores across all model configurations for both migration directions. The fine-tuned Gemma 3 12B achieves the highest scores in both directions, substantially outperforming the {\prevapproach} baseline (leftmost bars).}
        \label{fig:rq3_similarity-comparison}
    \end{figure}

\begin{figure}[t]
    \centering
    \includegraphics[width=.9\textwidth]{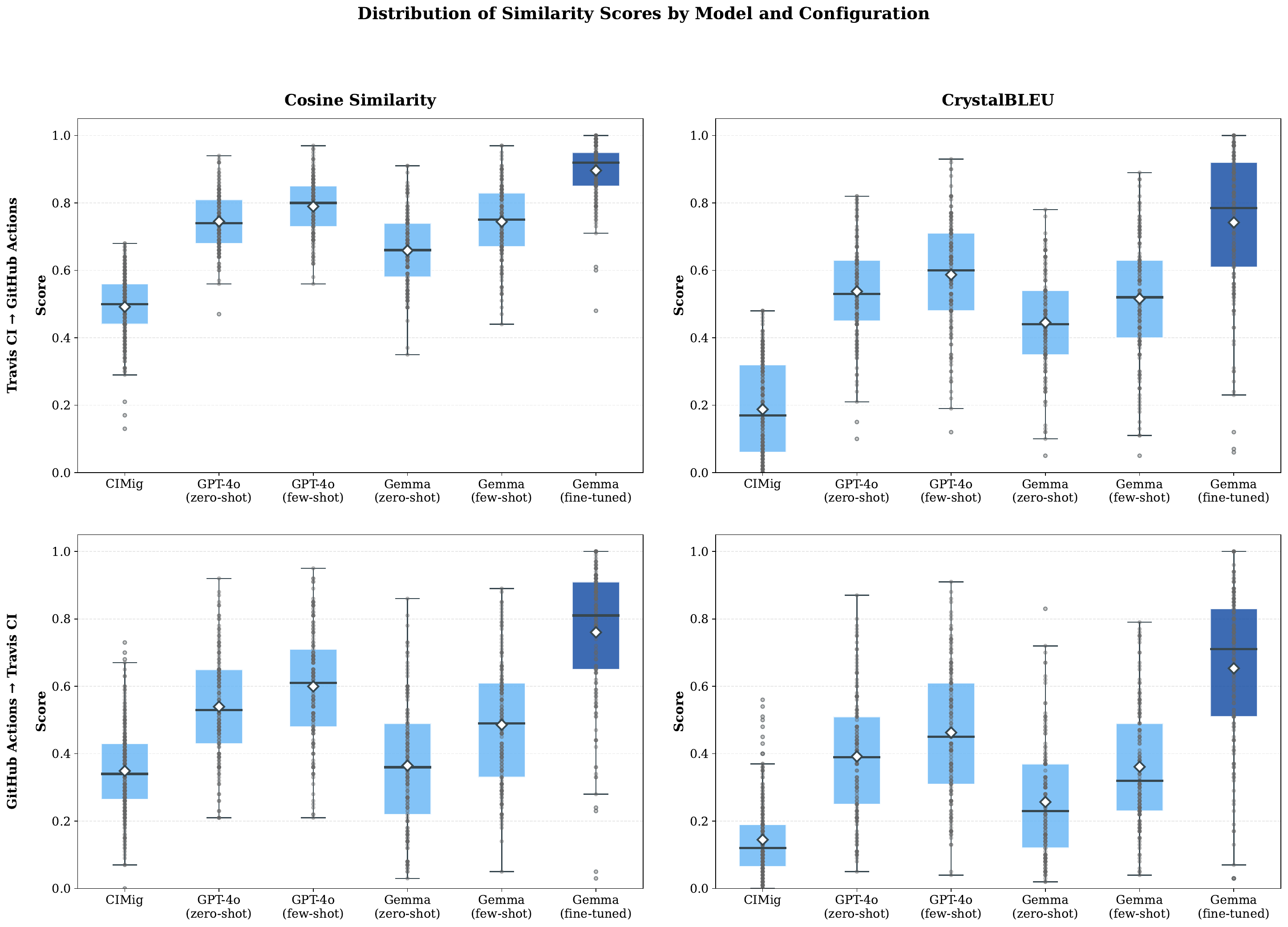}
    \caption{Distribution of similarity scores across model configurations. Box plots show median, interquartile range, and outliers. The fine-tuned Gemma model exhibits both higher median scores and lower variance compared to other configurations, indicating more consistent migration quality. {\prevapproach} (leftmost) shows substantial variance with many outliers.}
    \label{fig:boxplot-comparison}
    \end{figure}

    The fine-tuned Gemma~3 12B achieves the strongest overall performance over {\prevapproach}. On Travis$\rightarrow$GHA, it increases the mean per-project maximum similarity from 0.49 to 0.90 in Cosine (+82.2\%) and from 0.19 to 0.74 in CrystalBLEU (+295.5\%). On GHA$\rightarrow$Travis, it improves Cosine from 0.35 to 0.76 (+117.9\%) and CrystalBLEU from 0.14 to 0.65 (+351.2\%), all computed relative to the {\prevapproach} baseline on the same cross-validation projects. All improvements are statistically significant under a Mann-Whitney U test ($p < 10^{-48}$, $\delta \geq 0.87$, large effects). Figure~\ref{fig:rq3_similarity-comparison} provides a visual comparison of similarity scores across all model configurations and both migration directions, while Figure~\ref{fig:boxplot-comparison} shows the distribution of scores, revealing variance and outliers that mean values alone do not capture.
    
    \subsubsection{Syntactic Validity}
    
    Perhaps the most striking difference between {\approach} and {\prevapproach} lies in syntactic correctness. As detailed in Table~\ref{tab:linting-results}, {\prevapproach} produces valid YAML in only 5.6\% (Travis$\rightarrow$GHA) and 6.0\% (GHA$\rightarrow$Travis) of cases, with actionlint pass rates of 0\% for GHA outputs and travis-lint pass rates of only 6\% for Travis outputs. This suggests that the vast majority of {\prevapproach} outputs are not directly parseable as valid YAML and would likely require developer intervention before they could be used in practice.

    Automated error analysis of 502 {\prevapproach} outputs confirms that 94.2\% fail YAML parsing. The dominant failure mode is structural corruption (\textit{mapping values are not allowed here}: 68.5\% of failures), producing hybrid Travis/GHA syntax (e.g., misplaced \texttt{matrix:} blocks and malformed \texttt{jobs:} sections) that could necessitate substantial rework rather than minor edits. We did not measure repair time in this study; further investigation is needed to quantify the human effort required to bring invalid {\prevapproach} outputs to a lint-passing state.

    In contrast, {\approach}'s LLM-based outputs are overwhelmingly valid:
    
    \begin{itemize}[nosep,leftmargin=*]
        \item \textbf{Fine-tuned Gemma}: 100\% YAML validity and 89.3\% lint pass for Travis$\rightarrow$GHA; 97.3\% on both for GHA$\rightarrow$Travis
        \item \textbf{GPT-4o (few-shot)}: 99.4\% YAML validity and 90.9\% lint pass for Travis$\rightarrow$GHA; 96.1\% on both for GHA$\rightarrow$Travis
        \item \textbf{Gemma (zero-shot)}: 98.7\% YAML validity and 58.8\% lint pass for Travis$\rightarrow$GHA; 77.8\% on both for GHA$\rightarrow$Travis
    \end{itemize}

    Most LLM-based configurations significantly outperform CIMig on both YAML validity and CI linter pass rates according to two-proportion z-tests ($p < 0.001$, Bonferroni-corrected $\alpha = 0.0014$), with effect sizes ranging from medium to large (Cohen's $h$). The sole exception is LLaMA~3.1 8B in few-shot mode, which not only fails to improve but actively degrades YAML validity \emph{below} the CIMig baseline (Travis$\rightarrow$GHA: 2.0\% vs.\ 5.6\%; GHA$\rightarrow$Travis: 0.7\% vs.\ 6.0\%; $h < 0$). Among significant configurations, LLaMA~3.1 8B in zero-shot mode shows the smallest gains with medium effect sizes ($h \approx 0.6$), while Mistral 7B in zero-shot mode achieves large effect sizes ($h \approx 0.8$).
    
\begin{figure}[t]
        \centering
        \includegraphics[width=.9\textwidth]{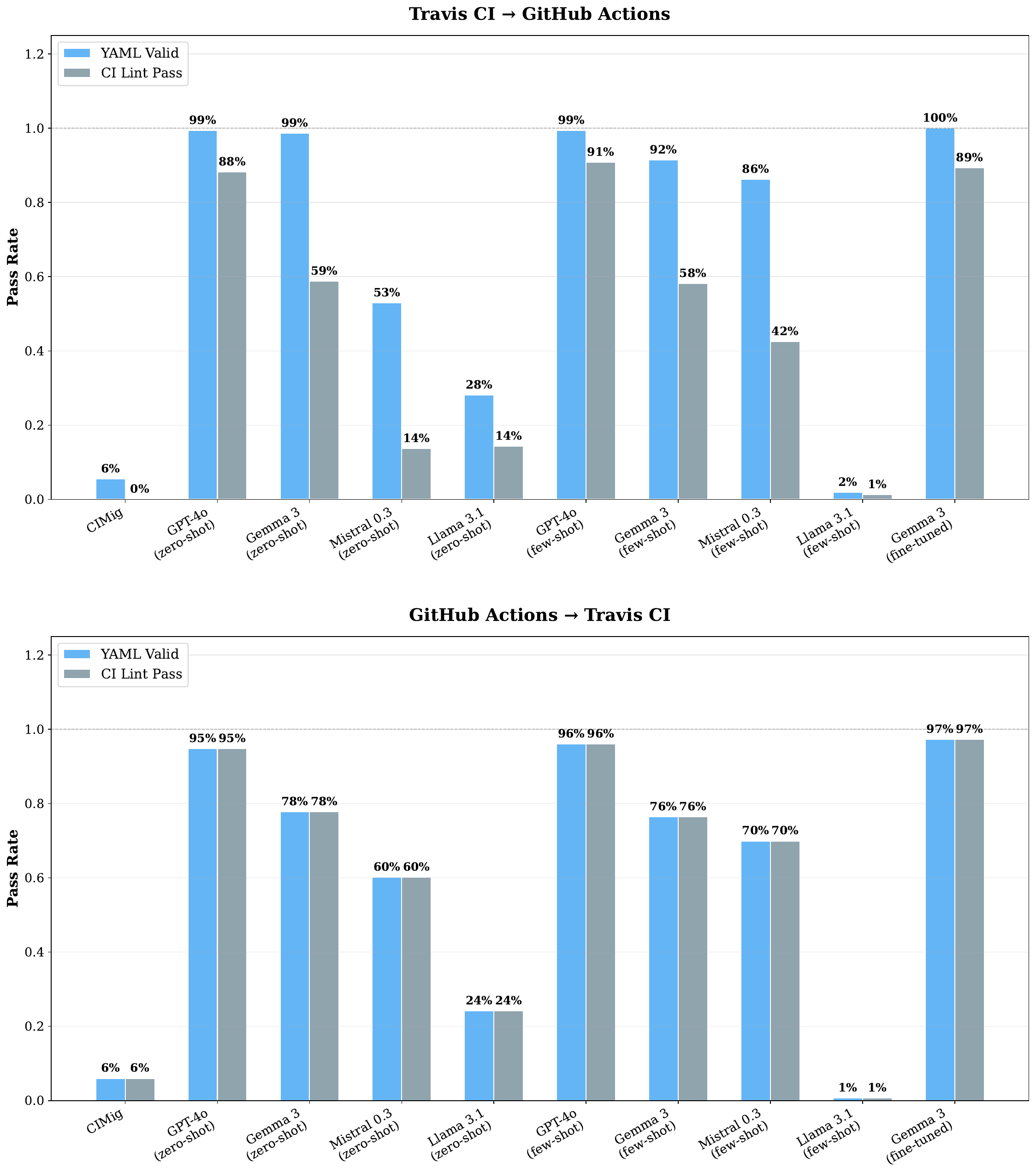}
        \caption{YAML validity and CI linter pass rates across all configurations. {\prevapproach} (leftmost) produces valid YAML in only 5.6\% to 6\% of cases with 0\% to 6\% lint pass rates. LLM-based approaches achieve dramatically higher validity, with the fine-tuned Gemma reaching 100\% YAML validity.}
        \label{fig:linting-results}
    \end{figure}

    This difference in syntactic correctness has major practical implications. {\approach}'s high validity rates mean its outputs can often be used directly or with minimal modification, whereas {\prevapproach}'s low parse and lint pass rates suggest that its outputs may function more as starting points that could require extensive manual debugging before deployment. Further work should validate the associated repair effort through controlled developer studies. Figure~\ref{fig:linting-results} shows the linting results across all configurations.
    
    \begin{table}[t]
    \centering
    \caption{Comprehensive comparison of {\approach} vs.\ {\prevapproach} across all configurations. Cosine Similarity and CrystalBLEU are reported for both migration directions. Bold indicates the best result per metric and direction.}
    \label{tab:rq3-comparison}
    \begin{tabular}{lcccc}
    \toprule
    \multirow{2}{*}{\textbf{Approach}} &
    \multicolumn{2}{c}{\textbf{Travis $\rightarrow$ GHA}} &
    \multicolumn{2}{c}{\textbf{GHA $\rightarrow$ Travis}} \\
    \cmidrule(lr){2-3} \cmidrule(lr){4-5}
    & \textit{Cosine} & \textit{CrystalBLEU}
    & \textit{Cosine} & \textit{CrystalBLEU} \\
    \midrule
    \textbf{{\prevapproach} (rule-based)} & 0.49 & 0.19 & 0.35 & 0.14 \\
    
    \midrule
    {\approach} (GPT-4o, zero-shot) & 0.74 & 0.54 & 0.54 & 0.39 \\
    {\approach} (GPT-4o, few-shot)  & 0.79 & 0.59 & 0.60 & 0.46 \\
    
    \midrule
    {\approach} (Gemma 3, zero-shot) & 0.66 & 0.44 & 0.37 & 0.26 \\
    {\approach} (Gemma 3, few-shot)  & 0.74 & 0.52 & 0.49 & 0.36 \\
    
    \midrule
    {\approach} (Mistral, zero-shot) & 0.64 & 0.33 & 0.31 & 0.16 \\
    {\approach} (Mistral, few-shot)  & 0.71 & 0.45 & 0.42 & 0.29 \\
    
    \midrule
    {\approach} (Llama, zero-shot) & 0.59 & 0.29 & 0.27 & 0.16 \\
    {\approach} (Llama, few-shot)  & 0.60 & 0.42 & 0.26 & 0.19 \\
    
    \midrule
    {\approach} (Gemma 3, fine-tuned) &
    \textbf{0.90} & \textbf{0.74} &
    \textbf{0.76} & \textbf{0.65} \\
    \bottomrule
    \end{tabular}
    \end{table}
    
    \begin{table}[t]
    \centering
    \caption{Syntactic validity comparison: YAML parsing success rate and CI-specific linter pass rate (actionlint for GHA, travis-lint for Travis CI). {\prevapproach} produces syntactically invalid outputs in most cases, while {\approach} achieves high validity rates.}
    \label{tab:linting-results}
    \begin{tabular}{llcccc}
    \toprule
    \textbf{Approach} & \textbf{Setting} &
    \multicolumn{2}{c}{\textbf{Travis $\rightarrow$ GHA}} &
    \multicolumn{2}{c}{\textbf{GHA $\rightarrow$ Travis}} \\
    \cmidrule(lr){3-4} \cmidrule(lr){5-6}
    & & \textit{PyYAML} & \textit{Actionlint}
    & \textit{PyYAML} & \textit{Travislint} \\
    \midrule
    {\prevapproach} & Rule-based & 5.6\% & 0.0\% & 6.0\% & 6.0\% \\
    \midrule
    GPT-4o & Zero-shot & 99.4\% & 88.2\% & 94.8\% & 94.8\% \\
    GPT-4o & Few-shot & 99.4\% & 90.9\% & 96.1\% & 96.1\% \\
    \midrule
    Gemma 3 12B & Zero-shot & 98.7\% & 58.8\% & 77.8\% & 77.8\% \\
    Gemma 3 12B & Few-shot & 91.5\% & 58.2\% & 76.5\% & 76.5\% \\
    Gemma 3 12B & Fine-tuned & \textbf{100\%} & \textbf{89.3\%} & \textbf{97.3\%} & \textbf{97.3\%} \\
    \midrule
    Mistral 7B & Zero-shot & 52.9\% & 13.7\% & 60.1\% & 60.1\% \\
    Mistral 7B & Few-shot & 86.3\% & 42.5\% & 69.9\% & 69.9\% \\
    \midrule
    Llama 3.1 8B & Zero-shot & 28.1\% & 14.4\% & 24.2\% & 24.2\% \\
    Llama 3.1 8B & Few-shot & 2.0\% & 1.3\% & 0.7\% & 0.7\% \\
    \bottomrule
    \end{tabular}
    \end{table}

    \subsubsection{Controlling for Data Quality}
    
    The results reported above use our curated 153-project dataset with manually verified ground truth. One might question whether {\approach}'s improvements stem from LLM capabilities or simply from higher-quality evaluation data. To isolate this factor, we evaluated {\approach} (zero-shot and few-shot only) on the original 251-project dataset using the same noisy ground truth as {\prevapproach}. We deliberately excluded fine-tuning from this comparison because training a model on noisy, semantically drifted ground truth would be counterproductive, it would teach the model to reproduce incorrect or misaligned migrations rather than faithful ones. Table~\ref{tab:noisy-comparison} reports these results.
    
    \begin{table}[t]
    \centering
    \caption{Performance comparison on the original 251-project dataset with noisy ground truth (same evaluation setup as {\prevapproach}). Bold indicates the best result per metric and direction.}
    \label{tab:noisy-comparison}
    \begin{tabular}{llcccc}
    \toprule
    \textbf{Approach} & \textbf{Setting} &
    \multicolumn{2}{c}{\textbf{Travis $\rightarrow$ GHA}} &
    \multicolumn{2}{c}{\textbf{GHA $\rightarrow$ Travis}} \\
    \cmidrule(lr){3-4} \cmidrule(lr){5-6}
    & & \textit{Cosine} & \textit{CrystalBLEU}
    & \textit{Cosine} & \textit{CrystalBLEU} \\
    \midrule
    \textbf{{\prevapproach}} & Rule-based & 0.49 & 0.18 & 0.35 & 0.13 \\
    \midrule
    GPT-4o & Zero-shot & 0.65 & 0.33 & 0.43 & 0.20 \\
    GPT-4o & Few-shot & \textbf{0.66} & \textbf{0.37} & \textbf{0.47} & \textbf{0.24} \\
    \midrule
    Gemma 3 12B & Zero-shot & 0.62 & 0.31 & 0.36 & 0.17 \\
    Gemma 3 12B & Few-shot & 0.61 & 0.31 & 0.36 & 0.18 \\
    \midrule
    Mistral 7B & Zero-shot & 0.61 & 0.26 & 0.31 & 0.14 \\
    Mistral 7B & Few-shot & 0.61 & 0.27 & 0.32 & 0.14 \\
    \midrule
    Llama 3.1 8B & Zero-shot & 0.62 & 0.29 & 0.33 & 0.15 \\
    Llama 3.1 8B & Few-shot & 0.63 & 0.32 & 0.34 & 0.19 \\
    \bottomrule
    \end{tabular}
    \end{table}
    
    Even on the original noisy dataset, all LLM configurations \emph{significantly} outperform {\prevapproach} on Travis$\rightarrow$GHA migration (all $p < 0.001$, Wilcoxon signed-rank test with Bonferroni correction), against a {\prevapproach} baseline of 0.49/0.18 (Cosine/CrystalBLEU). Effect sizes are mostly large ($r = 0.48$ to $0.97$): GPT-4o few-shot achieves the highest scores (0.66/0.37, $r = 0.96$/$r = 0.90$), while open-weight models also improve substantially, ranging from 0.61 to 0.63 in Cosine and 0.26 to 0.32 in CrystalBLEU (all $r \geq 0.48$, large or medium effects). The sole exception is Mistral~7B zero-shot CrystalBLEU ($r = 0.48$, medium), with all other comparisons showing large effects. For GHA$\rightarrow$Travis, only GPT-4o shows statistically significant improvements across both metrics against {\prevapproach}'s baseline of 0.35/0.13: zero-shot reaches 0.43/0.20 ($p < 0.001$, $r = 0.51$/$r = 0.51$, large effects) and few-shot extends this to 0.47/0.24 ($p < 0.001$, $r = 0.69$/$r = 0.72$, large effects). The consistent, significant advantage of LLM-based approaches on Travis$\rightarrow$GHA across both curated and noisy datasets confirms that performance improvements are attributable to LLM capabilities rather than data curation alone.
    
    \summarybox{\textbf{RQ3 Summary.}
    Compared to the rule-based {\prevapproach}, {\approach} consistently produces more accurate and more usable CI migrations. Fine-tuning achieves the strongest performance (Travis$\rightarrow$GHA: 0.90/0.74; GHA$\rightarrow$Travis: 0.76/0.65, both $p < 0.001$, $\delta \geq 0.87$). Prompting-only models also show substantial gains, with GPT-4o few-shot reaching 0.79/0.59 (Travis$\rightarrow$GHA) and 0.60/0.46 (GHA$\rightarrow$Travis) with high correlation to references ($r = 0.99/0.97$ and $r = 0.80/0.90$). In terms of syntactic validity, {\prevapproach} produces valid YAML in only 5.6\%/6.0\% of cases and passes CI linting in 0.0\%/6.0\%, whereas {\approach} reaches up to 100\%/97.3\% (fine-tuned) and 99.4\%/96.1\% (GPT-4o few-shot). These improvements persist even on the original 251-project noisy dataset, where all LLM variants significantly outperform {\prevapproach} on Travis$\rightarrow$GHA ($p < 0.001$), indicating gains stem from model capability rather than dataset curation.
    }   

\subsection{RQ4 Results}
\label{sec:rq4-results}
 
     While automated metrics provide valuable insights into migration accuracy, the ultimate measure of a migration tool's utility is whether developers find its outputs acceptable for real-world use. Table~\ref{tab:rq4-summary} presents the characteristics of the 30 pull requests submitted across both migration directions, highlighting migration success rates, collaboration effort, and developer engagement.

    \begin{table}[t]
    \centering
    \caption{RQ4: Migration success and engagement metrics for 30 submitted pull requests}
    \label{tab:rq4-summary}
    \begin{tabular}{lcc}
    \toprule
    \textbf{Metric} & \textbf{Travis→GHA} & \textbf{GHA→Travis} \\
    \midrule
    Total PRs Submitted & 15 & 15 \\
    \midrule
    \multicolumn{3}{l}{\textit{Migration Success (Cloud CI/CD Testing)}} \\
    Immediate Success (1 try) & 3 (20\%) & 6 (40\%) \\
    Minor Changes (2 tries) & 8 (53\%) & 6 (40\%) \\
    Small Changes (3 tries) & 2 (13\%) & 1 (7\%) \\
    Multiple Iterations (4+ tries) & 0 (0\%) & 1 (7\%) \\
    Failed & 2 (13\%) & 1 (7\%) \\
    \midrule
    \multicolumn{3}{l}{\textit{Collaboration Effort}} \\
    Sustained (2 follow-ups) & 10 (67\%) & 9 (60\%) \\
    Moderate (1 follow-up) & 5 (33\%) & 4 (27\%) \\
    \midrule
    \multicolumn{3}{l}{\textit{Developer Feedback}} \\
    With Reviews & 5 (33\%) & 6 (40\%) \\
    With Explicit Ratings & 1 (7\%) & 0 (0\%) \\
    PR Status (Open/Closed/Merged) & 9/5/1 & 4/11/0 \\
    \bottomrule
    \end{tabular}
    \end{table}

    \subsubsection{Migration Success Rates}

    \paragraph{Immediate Success Rate.}
    A critical finding is that 30\% of the selected migrations (9 out of 30 PRs) succeeded immediately without any modifications, the generated configuration passed all checks and executed successfully in the cloud CI environment on the first attempt. When comparing migration directions, the immediate success rate was notably higher for GHA$\rightarrow$Travis migrations (40\%) compared to Travis$\rightarrow$GHA migrations (20\%), suggesting that the fine-tuned model handles the compression task (GHA's verbose syntax to Travis's compact format) more reliably than the expansion task.

    \paragraph{Minor Refinement Requirements.}
    The majority of migrations (47\%, 14 out of 30) required only minor changes, typically 2 attempts to achieve a passing build. These refinements addressed issues such as:

    \begin{itemize}[nosep,leftmargin=*]
        \item Environment variable naming mismatches
        \item Project-specific path corrections
        \item Dependency version specifications
        \item Cache key formatting
    \end{itemize}

    Combined, 77\% of migrations succeeded with at most 2 attempts (23 out of 30), demonstrating that {\approach} produces configurations that are either immediately usable or require only trivial adjustments.

    \paragraph{Complex Cases Requiring Multiple Iterations or Failing.}
    The remaining 23\% of migrations (7 out of 30) required three or more attempts or failed entirely. This includes three migrations that succeeded after three attempts, one that required four or more iterations, and three that did not converge during our evaluation. These cases involved project-specific complexities such as:

    \begin{itemize}[nosep,leftmargin=*]
        \item Custom build matrix configurations with non-standard language versions
        \item Complex deployment pipelines with external secret management
        \item Travis CI log size limits (4MB) requiring output suppression
        \item GitHub Actions-specific features without Travis equivalents (e.g., reusable workflows)
    \end{itemize}

    \paragraph{Real-World Adoption: A Merged Pull Request.}
    The strongest evidence of practical deployment acceptance is that one Travis$\rightarrow$GHA pull request was merged into the upstream repository: {NLPchina/elasticsearch-sql}. This migration was submitted to an actively maintained project, passed the repository's CI checks, and was accepted by the project maintainers without requesting structural changes. A merged PR represents the full deployment lifecycle: the generated configuration executed successfully in the real CI environment \emph{and} was deemed production-ready by the project team. While a single merged PR is a limited sample, it demonstrates that {\approach} can produce configurations that meet the quality bar for real open-source projects.

    \subsubsection{Developer Feedback and Deployment Context}

    Table~\ref{tab:rq4-feedback} presents the distribution of feedback categories across all 30 PRs.

    \begin{table}[t]
    \centering
    \caption{RQ4: Categorization of developer feedback across 30 pull requests}
    \label{tab:rq4-feedback}
    \begin{tabular}{lcc}
    \toprule
    \textbf{Feedback Category} & \textbf{Count} & \textbf{Percentage} \\
    \midrule
    No Response Yet & 19 & 63.3\% \\
    Response Received & 7 & 23.3\% \\
    Not Interested & 3 & 10.0\% \\
    Technical Issues Identified & 1 & 3.3\% \\
    \bottomrule
    \end{tabular}
    \end{table}

    \paragraph{Response Received (23.3\%).}
    Seven PRs received developer feedback, with one project (Umweltzone/Umweltzone) providing an explicit 4-star rating (out of 5). The developer noted: ``The conversion result looks very good in general. Also, your test build succeeded as you stated.'' This PR required 3 attempts to achieve success and involved sustained collaboration (1 follow-up). Other feedback highlighted that generated configurations captured build matrices, caching strategies, and deployment logic with minimal manual intervention.

    \paragraph{Technical Issues (3.3\%).}
    One PR received feedback identifying specific technical issues requiring correction, such as Java version mismatches and missing Git LFS handling. Notably, this feedback came from a project (jonasoreland/runnerup) where the migration ultimately required 3 attempts before achieving a passing build, demonstrating that {\approach} can handle iterative refinement based on developer input.

    \paragraph{Not Interested (10\%).}
    Three maintainers indicated that they were not interested in the proposed CI change at the time, or that they would revisit it later. Importantly, these responses did not critique the migration quality itself; instead, they reflected project-specific priorities and timing constraints:

    \begin{itemize}[nosep,leftmargin=*]
        \item ``This is a dying project after spring has introduced their own approach. I will not use time on it.'' (ksaua/remock)
        \item ``I don't plan to migrate to GH, but if I decide to, I will check this PR, thanks.'' (javers/javers)
        \item ``Thank you for your enquiry. We will look to review this in the new year and feedback.'' (ikasanEIP/ikasan)
    \end{itemize}

    These responses suggest that rejection was due to project-specific circumstances (project abandonment, lack of immediate migration plans) rather than inadequate migration quality.

    \paragraph{No Response (63.3\%).}
    The majority of PRs (19 out of 30) have not yet received developer responses. This aligns with prior studies on open-source contribution response rates~\cite{yu2015wait}, where unsolicited pull requests often experience delayed or no responses, particularly in less actively maintained projects. The sustained collaboration effort reflected in our selection (67\% of Travis$\rightarrow$GHA PRs had 2 follow-ups) demonstrates proactive researcher engagement to encourage feedback.

    \summarybox{\textbf{RQ4 Summary.}
    Evaluated across 30 real-world pull requests (15 per direction), {\approach} shows strong practical deployment performance. 9/30 (30\%) configurations pass CI checks on the first attempt, and 23/30 (77\%) succeed within two attempts, with most fixes involving minor project-specific adjustments (paths, versions, cache keys, environment variables). The remaining 7/30 (23\%) require three or more attempts or do not converge, typically in projects with more complex platform-specific CI setups. Notably, one Travis$\rightarrow$GHA pull request was merged upstream, demonstrating end-to-end adoption. Maintainer feedback was limited overall (19/30 no response); among responses, 4/30 (13\%) were positive and 3/30 (10\%) cited project constraints unrelated to migration quality. CI execution success is therefore the primary deployment signal, with maintainer responses providing supplementary evidence of usability.
    }

\section{Discussion}
\label{sec:discussion}

This section discusses the key findings from our evaluation and their practical implications for CI migration practice.

\vspace{3pt}
\noindent \textit{\textbf{LLMs fundamentally outperform rule-based approaches for CI migration.}} Our results demonstrate that LLM-based approaches substantially outperform the rule-based {\prevapproach} across virtually all configurations and metrics. Even the simplest LLM setup (zero-shot prompting with an 8B parameter model) achieves higher similarity scores than {\prevapproach} in most cases. This finding is consistent with LLMs' broad code-generation capabilities acquired through pre-training, and the advantage persists on both our curated 153-project set and the original 251-project noisy corpus (Table~\ref{tab:noisy-comparison}). Because source configurations originate from public repositories, we cannot exclude that pre-training exposure contributes; our expert-authored ground truth mitigates reference memorization, but post-cutoff evaluation remains future work.

\vspace{3pt}
\noindent \textit{\textbf{Syntactic validity is the critical differentiator.}} While similarity metrics show meaningful improvements, the most practically significant finding is the dramatic difference in syntactic validity. {\prevapproach} produces valid YAML in only 5.6\% to 6\% of cases; automated error analysis shows that 68.5\% of parse failures involve structural corruption (\textit{mapping values are not allowed here}), which could require reorganization rather than small edits. In contrast, the fine-tuned Gemma model achieves 100\% YAML validity for Travis$\rightarrow$GHA and 97.3\% for GHA$\rightarrow$Travis. This difference fundamentally changes the user experience: {\approach}'s outputs can often be used directly or with minimal modification, whereas {\prevapproach}'s outputs would likely require substantial manual correction before they could be put into production. Further investigation should quantify the repair effort associated with invalid {\prevapproach} outputs to confirm this practical gap.

\vspace{3pt}
\noindent \textit{\textbf{Migration direction asymmetry reflects inherent task difficulty.}} We consistently observed that GHA$\rightarrow$Travis migration is more challenging than Travis$\rightarrow$GHA, with lower similarity scores and requiring different hyperparameters during fine-tuning. Fine-tuned GHA$\rightarrow$Travis exhibits 2.2$\times$ higher per-project cosine variance than Travis$\rightarrow$GHA ($\sigma = 0.20$ vs.\ $0.09$) and wider fold-level dispersion (fold means 0.63--0.85 vs.\ 0.86--0.94). The weakest fold was driven by projects with multi-job workflows and build matrices (e.g., \texttt{vert-x3\_vertx-bridge-common}, \texttt{OpenLiberty\_ci.maven}), supporting the view that compressing explicit GHA structure into Travis's compact format is inherently harder than the reverse expansion.

\vspace{3pt}
\noindent \textit{\textbf{Few-shot prompting provides consistent but modest improvements.}} Across most models, few-shot prompting improved performance over zero-shot by 5\% to 15\% on similarity metrics. This improvement comes essentially ``for free'', requiring only the addition of a single example to the prompt without any model training. However, we also observed that few-shot prompting can backfire for smaller models: Llama~3.1 8B few-shot validity fell to 0.7--2.0\% because 95\% of failures reproduced markdown-wrapped example formatting rather than pure YAML, despite CrystalBLEU gains ($p < 0.001$). Practitioners should therefore validate that few-shot examples improve both similarity and syntactic validity for their specific model before adopting them in production.

\vspace{3pt}
\noindent \textit{\textbf{Fine-tuning offers the best performance.}} The fine-tuned Gemma model achieved the highest scores across all metrics, demonstrating that domain-specific adaptation provides meaningful benefits beyond prompting strategies alone. However, fine-tuning requires curating a training dataset, which involved substantial manual effort in our case (creating 153 ground-truth migration pairs with expert developers). For occasional migration needs ($<$10 projects), GPT-4o few-shot is more practical (about 14\,s per project; about 35\,min per 153-project batch). Using \texttt{tiktoken} on the actual prompts and saved outputs, a full 153-project GPT-4o batch costs \$0.57--\$0.73 per direction (mean 622 input and 306--319 output tokens for Travis$\rightarrow$GHA; 782 input and 176--212 output tokens for GHA$\rightarrow$Travis at OpenAI list rates of \$2.50/\$10.00 per million input/output tokens). Fine-tuning both directions required a one-time cost of approximately 62 to 68 minutes (31 to 35 minutes per direction, including model loading, training, evaluation, and saving; Table~\ref{tab:efficiency-comparison}), comparable to the wall-clock time of roughly two 153-project GPT-4o batches (about 35 minutes each). After this one-time cost, further migrations require no additional training, only inference (about 14 seconds per project locally), whereas each additional GPT-4o batch continues to incur both time and API cost. This favors fine-tuning once ground-truth curation is already sunk and migration volume is recurring. An alternative to full fine-tuning is Retrieval-Augmented Generation (RAG)~\cite{merritt2025rag}, which could dynamically retrieve relevant migration examples or CI service documentation at inference time, potentially offering similar benefits without the overhead of model training.

\vspace{3pt}
\noindent \textit{\textbf{Efficiency trade-offs favor LLM-based approaches for practical use.}} Table~\ref{tab:efficiency-comparison} compares the computational costs of {\approach} and {\prevapproach}. While {\prevapproach}'s rule-based inference is faster (0.7 to 1.2 seconds per migration vs.\ 24 to 38 seconds for {\approach}), its preprocessing phase is substantially more expensive: Frequent-Tree Mining alone requires approximately 71 hours on GitHub Actions files, plus an additional hour for Travis CI mining, rule matching, and Apriori operations, totaling roughly 72 hours of preprocessing. In contrast, fine-tuning {\approach} requires only 31 to 35 minutes per direction (including model loading, training, evaluation, and saving), making it approximately 130$\times$ faster to prepare. When normalized by dataset size, {\prevapproach} requires about 9 seconds per file for preprocessing (on about 29,000 files), while {\approach} requires about 14 seconds per project for fine-tuning (on 138 projects), comparable per-item costs, but with {\approach} achieving dramatically better accuracy. Moreover, the inference time difference is negligible in practice: CI migration is typically a one-time task per project, and the additional 20 to 35 seconds is modest relative to the remediation effort that {\prevapproach}'s 94\% YAML failure rate and structural error profile would likely entail---though this should be confirmed in future work. On the {\approach} side, RQ4 shows a median of two cloud-CI iterations to reach a passing build (mean Success Try \# = 1.9), providing a proxy for post-generation effort. From a practical standpoint, {\approach} offers a more favorable trade-off: substantially lower preparation costs, comparable per-item efficiency, and substantially higher syntactic validity.

\begin{table}[t]
\centering
\caption{Efficiency comparison between {\approach} (fine-tuned) and {\prevapproach}}
\label{tab:efficiency-comparison}
\begin{tabular}{lcc}
\toprule
\textbf{Metric} & \textbf{{\prevapproach}} & \textbf{{\approach} (Fine-tuned)} \\
\midrule
Preprocessing/Training time & 72 hours & 31--35 minutes \\
Training data size & 29,000 files & 138 projects \\
Normalized prep.\ cost & 9 sec/file & 14 sec/project \\
\midrule
Inference (Travis$\rightarrow$GHA) & 0.7 sec & 38 sec \\
Inference (GHA$\rightarrow$Travis) & 1.2 sec & 24 sec \\
\midrule
Cosine Similarity (Travis$\rightarrow$GHA) & 0.49 & 0.90 \\
YAML Validity (Travis$\rightarrow$GHA) & 5.6\% & 100\% \\
\bottomrule
\end{tabular}
\end{table}

\section{Threats To Validity}
\label{sec:threats_to_validity}
\vspace{-2pt}

\subsection{Construct Validity}
\vspace{-2pt}
Construct validity concerns whether our metrics accurately measure what we intend to measure. We use Cosine Similarity and CrystalBLEU to assess semantic similarity between generated and reference configurations. While these metrics are established in code generation evaluation~\cite{eghbali2022crystalbleu,gunawan2018cosine}, they measure textual similarity rather than functional equivalence. A configuration with different syntax but identical behavior would score lower than warranted, while a configuration with similar text but different semantics might score higher. To mitigate this threat, we complement similarity metrics with linting validation, which provides a functional check that configurations can at least be parsed and validated by the target CI service. Our RQ4 developer study provides human assessment of configuration quality beyond automated metrics.

Our linting evaluation uses standard tools (actionlint, travis-lint) that may not catch all semantic issues. A configuration might pass linting but still fail during actual CI execution due to missing secrets, incorrect paths, or runtime dependencies. We acknowledge this limitation and note that our developer study (RQ4) provides a more comprehensive assessment of practical usability.

\subsection{Internal Validity}
\vspace{-2pt}

Internal validity concerns whether our experimental design supports valid conclusions about the relationship between LLM configurations and migration quality. One threat is potential data leakage: the evaluated LLMs may have seen CI configurations from our evaluation projects during pre-training. To mitigate this, we manually curated a ground-truth dataset in which two expert developers created new GitHub Actions workflows for each Travis CI configuration, ensuring the reference outputs are novel rather than potentially memorized.

Another threat involves hyperparameter selection. Our fine-tuning experiments required tuning learning rates separately for each migration direction, and different choices might yield different results. We addressed this by performing systematic hyperparameter search and reporting the process transparently. The Wilcoxon signed-rank test we use for statistical comparisons is robust to distribution assumptions, providing reliable significance testing.

\subsection{External Validity}
\vspace{-2pt}
External validity concerns the generalizability of our findings. Our evaluation uses 153 projects from the {\prevapproach} dataset, which may not represent the full diversity of CI configurations in practice. The projects are predominantly Java-based and may not reflect configurations for other languages or more complex multi-language projects. Similarly, we evaluate only Travis CI and GitHub Actions; our findings may not generalize to other CI services (e.g., CircleCI, GitLab CI, Jenkins).
In addition, our choice of LLMs (GPT-4o, Gemma 3, Mistral, Llama) represents a snapshot of available models as of our evaluation. LLM capabilities evolve rapidly, and newer models may perform differently. We mitigate this by using both proprietary and open-weight models of various sizes, providing a broader view of LLM capabilities for this task.

\subsection{Conclusion Validity}
\vspace{-3pt}

Conclusion validity concerns whether our statistical conclusions are justified. We use the Wilcoxon signed-rank test with a standard significance threshold ($\alpha = 0.05$) and report exact p-values. The extremely low p-values we observe ($p < 10^{-26}$ for key comparisons) provide strong evidence for our conclusions, making Type I errors unlikely. However, statistical significance does not imply practical significance; we therefore complement our statistical tests with effect size reporting.

The sample size difference between zero/few-shot experiments ($n=153$) and fine-tuning experiments ($n=150$) arises from the 10-fold cross-validation design requiring divisible fold sizes. This slight difference does not materially affect our conclusions, as both sample sizes provide adequate statistical power.

Our evaluation uses a manually curated 153-project dataset (Section~\ref{sec:dataset}) rather than the full 1,252-project {\prevapproach} corpus. Although smaller, the sample size was determined through a formal sample-size calculation, and manual curation substantially improved ground-truth reliability by eliminating semantically drifted workflow pairs. While a larger curated dataset would strengthen confidence in the magnitude of the observed effects, we do not expect it to change their overall direction.

For RQ4, developer evaluation was limited by the low response rate to unsolicited pull requests. As noted in Section~\ref{sec:pull_requests}, only one of 30 PRs included an explicit Likert-scale rating, precluding meaningful inferential analysis. We therefore report developer ratings and comments descriptively rather than statistically. Consequently, the RQ4 findings should be interpreted as suggestive rather than statistically confirmed.

\section{Conclusion}
\label{sec:conclusion}
\vspace{-3pt}

This paper presented {\approach}, an LLM-based framework for automating CI configuration migration between Travis CI and GitHub Actions. Through an empirical evaluation comparing four LLMs under zero-shot, few-shot, and fine-tuned settings against the rule-based {\prevapproach} baseline, we demonstrated that LLM-based approaches substantially outperform traditional rule-based techniques for CI migration.

Our key findings include the following. 
\vspace{-2pt}

\begin{itemize}
\item \textbf{Zero-shot performance:} Even without task-specific training, zero-shot LLMs significantly outperform {\prevapproach} on similarity metrics. GPT-4o zero-shot achieves 0.74 Cosine Similarity and 0.54 CrystalBLEU for Travis$\rightarrow$GHA migrations (+51\% and +184\% over the baseline), while few-shot prompting further improves GPT-4o to 0.79/0.59 (+61\% and +211\% over the baseline).

\item \textbf{Fine-tuning gains:} Fine-tuning the best open-weight model (Gemma 3 12B) on a curated dataset further improves performance to 0.90/0.74 for Travis$\rightarrow$GHA and 0.76/0.65 for GHA$\rightarrow$Travis, corresponding to improvements of 82.2\% to 117.9\% in Cosine Similarity and 295.5\% to 351.2\% in CrystalBLEU over {\prevapproach}.

\item \textbf{Practical usability:} {\approach} produces syntactically valid YAML in 89\% to 100\% of cases compared to only 5.6\% to 6\% for {\prevapproach}, with CI linter pass rates of 58\% to 97\% versus 0\% to 6\%. This indicates that {\approach} outputs are more immediately usable in practice, whereas {\prevapproach} outputs would likely require extensive manual correction, thus warranting further empirical validation.
\end{itemize}

These results suggest that LLMs offer a more practical and accurate solution for CI configuration migration than rule-based approaches. For practitioners, we recommend using GPT-4o with few-shot prompting for occasional migration needs, or investing in fine-tuning an open-weight model like Gemma for organizations with frequent migration requirements. For researchers, our work opens avenues for exploring LLM-based automation of other DevOps configuration tasks, such as Dockerfile generation, Kubernetes manifest migration, or infrastructure-as-code transformations.

\medskip\noindent\textbf{Future work.} We aim to extend {\approach} to support additional CI services (e.g., CircleCI, GitLab CI, Jenkins), investigate multi-step migration refinement using LLM agents, explore Retrieval-Augmented Generation (RAG) to fetch relevant migration examples and CI documentation at inference time, develop methods to ensure functional equivalence beyond syntactic correctness, and measure the human effort required to repair invalid {\prevapproach} outputs through controlled developer studies.

\section*{Declarations}
\vspace{-2pt}

\bmhead{Funding} This research was supported by the Natural Sciences and Engineering Research Council (NSERC) Discovery Grant [RGPIN-2025-05897].

\bmhead{Acknowledgments}
The experiments conducted in this paper were enabled in part by support provided by the Digital Research Alliance of Canada. 

\bmhead{Author Contributions}{
Md Nazmul Hossain: Conceptualization, Investigation, Formal analysis, and Writing. 
Taher A. Ghaleb: Conceptualization, Investigation, Formal analysis, and Writing.
}

\bmhead{Data availability Statement}
The replication package for our experiments, including data, code, and results, is available on GitHub~\cite{our_replication_package}.

\bmhead{Conflicts of Interest} The authors declare that they have no known competing financial interests or personal relationships that could have appeared to influence the work reported in this paper.

\bmhead{Ethical approval} Not applicable.

\bmhead{Informed consent} Not applicable. 

\bmhead{Clinical Trial Number} Not applicable.

\bibstyle{sn-mathphys-num}
\bibliography{paper}

\end{document}